%% file: flash_channel_modeling_journal.tex
\documentclass[journal, twoside]{IEEEtran}
\usepackage{cite}

\ifCLASSINFOpdf
\else
\fi
\usepackage[cmex10]{amsmath}
\usepackage{centernot}
\usepackage{graphics}
\usepackage{tikz}
\usepackage{pgfplots}
\usetikzlibrary{pgfplots.groupplots, patterns, decorations.pathreplacing}

\newtheorem{IEEEprop}{Proposition}

\newtheorem{IEEEdefinition}{Definition}

\usepackage{bm}

\DeclareMathOperator{\E}{E}
\DeclareMathOperator{\Var}{Var}

\usepackage{algorithm}
\usepackage[noend]{algorithmic}

\usepackage{multirow}
\usepackage{hhline}

\usepackage{subfiles}

\newcommand{\tr}[1] {\textcolor{red}{#1}}
\newcommand{\tb}[1] {\textcolor{blue}{#1}}
\newcommand{\ca}{{\sim}}
\newcommand{\kcount}[2]{\ensuremath{K_{#1}^{(#2)}}}
\newcommand{\kcountm}[1]{\ensuremath{K_{#1}}}
\newcommand{\ZOerror} {\mbox{\ensuremath{0 \rightarrow 1}}}
\newcommand{\OZerror} {\mbox{\ensuremath{1 \rightarrow 0}}}
\newcommand{\indep}{\mathrel\bot\joinrel\mspace{-8mu}\mathrel\bot}

\pgfmathdeclarefunction{gauss}{2}{%
	\pgfmathparse{1/(#2*sqrt(2*pi))*exp(-((x-#1)^2)/(2*#2^2))}%
}

\begin{document}
\bstctlcite{IEEEexample:BSTcontrol}
%
\title{Channel Models for Multi-Level Cell Flash Memories Based on Empirical Error Analysis}
%
%
%

\author{Veeresh Taranalli,~\IEEEmembership{Student Member, IEEE,}
        Hironori Uchikawa,~\IEEEmembership{Member, IEEE,}
        Paul H. Siegel,~\IEEEmembership{Fellow, IEEE}
\thanks{This work was supported by National Science Foundation (NSF) Grants CCF-1116739, CCF-1405119 and the Center for Memory and Recording Research, UC San Diego. The material in this paper was presented in part at the IEEE International Conference on Communications, London, UK, June~8-12,~2015 and the Annual Non-Volatile Memories Workshop (NVMW), San Diego, USA, March~6-8,~2016.}
\thanks{V. Taranalli, and P. H. Siegel are with the University of California, San Diego, La Jolla, CA 92093-0401, USA (e-mail: {vtaranalli, psiegel}@ucsd.edu).}
\thanks{H. Uchikawa is with Toshiba Corporation, Japan (e-mail: hironori.uchikawa@toshiba.co.jp).}}

\maketitle
%
%
\begin{abstract}
We propose binary discrete parametric channel models for multi-level cell (MLC) flash memories that provide accurate ECC
performance estimation by modeling the empirically observed error characteristics under program/erase \mbox{(P/E)} cycling stress. Through a detailed empirical error characterization of \mbox{1X-nm} and \mbox{2Y-nm} MLC flash memory chips from two different vendors, we observe and characterize the \emph{overdispersion} phenomenon in the number of bit errors per ECC frame. A well studied
channel model such as the binary asymmetric channel (BAC) model is unable to provide accurate ECC
performance estimation. Hence we propose a channel model based on the beta-binomial probability distribution (\mbox{2-BBM} channel model) which is a good fit for the overdispersed empirical error characteristics and show through statistical tests and simulation results for BCH, LDPC and polar codes, that the \mbox{2-BBM} channel model provides accurate ECC
performance estimation in MLC flash memories.
%
%
\end{abstract}
\begin{IEEEkeywords}
	Flash memory, multi-level cell, channel model, error correcting codes, P/E cycling.
\end{IEEEkeywords}

%
\IEEEpeerreviewmaketitle

\subfile{sections/introduction}

%

\subfile{sections/flash_memory_structure}

%
\subfile{sections/experiment_procedure}
%

\subfile{sections/error_characterization}

%

\subfile{sections/dmc_for_flash}

%

\subfile{sections/ecc_performance_prediction}

%

\subfile{sections/conclusion}
\ifCLASSOPTIONonecolumn

\subfile{sections/appendix_singlecol}

\else

\subfile{sections/appendix}

\fi




%





\ifCLASSOPTIONcaptionsoff
  \newpage
\fi




\bibliographystyle{IEEEtran}
\bibliography{flash_channel_modeling_journal}
\end{document}

%% file: sections/introduction.tex
\section{Introduction}
\label{sec:introduction}

\IEEEPARstart{C}{hannel} modeling for NAND flash memories is a developing research area with applications to better signal processing and coding techniques. A channel model for a flash memory can be viewed as a simplified representation of the underlying physical mechanisms which induce errors in stored data. For NAND flash memories, the major error mechanisms are program disturb and cell wear that occur during program/erase cycling, charge loss that occurs during data retention and inter-cell interference (ICI)~\cite{Bez_2003, Cooke_2007, Lee_2002}. The main applications of a flash memory channel model are improved design, decoding and performance evaluation of error-correcting codes (ECCs) and error-mitigating codes. Other applications include information theoretic studies that provide an analysis of the capacity of flash memories~\cite{Huang_2013}, as well as insights for the development of new coding techniques. In this paper, we focus on the development of parametric channel models for multi-level cell (MLC) flash memories based on empirical error characterization, that enable accurate ECC frame error rate (FER) performance estimation/prediction.

\subsection{Overview of the Problem}
Efficient evaluation of ECC FER performance is important for storage system design and optimization. One approach to ECC FER performance estimation is to experimentally collect error data for use in Monte-Carlo simulations of the ECC decoder, but this can be impractical because of the large amount of error data required when estimating low frame error rates. Another approach is to analytically predict the performance of a code based upon a measured average raw bit error rate. While this is feasible for algebraic codes with bounded distance decoders, it is difficult for low density parity check (LDPC) codes and polar codes that use probabilistic decoders based upon message passing or successive cancellation. Moreover, the implicit assumption of independent, symmetric bit errors may not be justified.

Previously proposed~\cite{Cai_2013, Parnell_2014} parametric channel models for MLC flash memories were obtained by using well known probability distributions to model the empirical cell threshold voltage distributions. In~\cite{Cai_2013}, a Gaussian distribution, and in~\cite{Parnell_2014}, a Normal-Laplace mixture model were shown to be a good fit for the experimentally observed cell threshold voltage distributions in MLC flash memories.
Such models can be used to reliably predict/estimate the experimentally observed raw bit error rate (RBER) of the flash memory.
However in this paper, we show through empirical error characterization that the RBER is not necessarily a good indicator of the ECC FER performance and this is due to the overdispersion phenomenon in the number of bit errors per frame in MLC flash memories. Overdispersion refers to the greater variability in empirical data compared to a statistical model for e.g., the binomial distribution typically used to model count data. Therefore, a memoryless channel model such as the binary asymmetric channel (BAC) model provides an optimistic estimate of the ECC FER performance when compared to the actual ECC FER performance estimate obtained from empirical data.


\subsection{Summary of Contributions}
We present a detailed empirical characterization of errors in MLC flash memories at the bit, cell and page granularity levels for \mbox{1X-nm} and \mbox{2Y-nm} feature size MLC flash memory chips from two different vendors referred to as \mbox{vendor-A} and \mbox{vendor-B} respectively. We study the asymmetry of bit errors in the lower and upper pages of MLC flash memories with a focus on the \emph{number of bit errors per frame} parameter. We observe that the empirical probability distributions of the number of bit errors per frame parameter are \emph{overdispersed} when compared to a binomial distribution typically used to model count data.

Based on the empirical error analysis, we study the per-page binary asymmetric channel (BAC) model referred to as the 2-BAC model for MLC flash memories. Using statistical analysis, we show that the 2-BAC model does not provide a good fit for the empirical error data and hence is inadequate for accurate ECC frame error rate (FER) performance estimation. Therefore, we propose a channel model based on the beta-binomial probability distribution referred to as the 2-Beta-Binomial (2-BBM) channel model. We show that it is a good fit for the observed overdispersed empirical error data and performs well for ECC FER performance estimation. We also propose normal and Poisson approximation based channel models for MLC flash memories.

Through quantitative evaluation of the proposed channel models using the statistical Kolmogorov-Smirnov \mbox{(K-S)} Two Sample goodness of fit test and using Monte-Carlo simulation results of FER performance for BCH, LDPC and polar codes, we show that the 2-Beta-Binomial channel model is an accurate channel model to represent the overdispersed nature of bit errors in MLC flash memories.
\subsection{Organization of the Paper}
The rest of the paper is organized as follows. Section~\ref{sec:flash_memory_structure} presents a brief introduction to flash memories with a focus on the structure of MLC flash memories. In Section~\ref{sec:expt_proc} we describe the P/E cycling experiment procedure. Section~\ref{sec:cell_err_cap_ch} provides a detailed empirical characterization of errors in MLC flash memories, the results of which are utilized for design and evaluation of the proposed channel models. Section~\ref{sec: dmc_for_mlc_flash} describes the proposed channel models for MLC flash memories and provides statistical analysis results. In Section~\ref{sec:empirical_results}, quantitative results for statistical goodness of fit tests and BCH, LDPC and polar code FER performance are presented to evaluate the proposed channel models. Section~\ref{sec:conclusion} provides the concluding remarks.

%% file: sections/flash_memory_structure.tex
\section{Flash Memory Structure}
\label{sec:flash_memory_structure}
\begin{figure}
    \centering
    \input{tikz/flash_memory_block.tex}
    \caption{Cell level to bit mapping and block schematic in MLC flash memories. In the block schematic, the rectangles depict the MLC flash memory cells connected to horizontal wordlines (WL) and vertical bitlines (BL).} \label{fig:flash_block}
\end{figure}
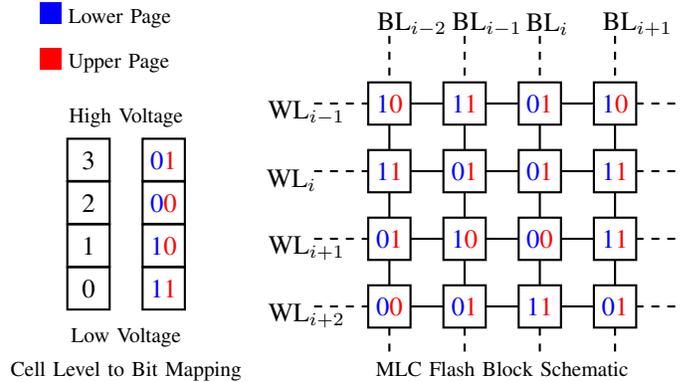

The fundamental data storing unit in NAND flash memories is a floating-gate transistor commonly referred to as a cell. A cell can be programmed to hold different levels of charge and these charge levels represent the data bits stored in a cell. The most commonly used cells in today's flash memories are capable of holding 2, 4 and 8 distinct charge levels (1, 2, 3 bits/cell respectively) and are referred to as single-level cell (SLC), multi-level cell (MLC) and three-level cell (TLC) respectively. These flash memory cells are organized into a rectangular array interconnected through horizontal wordlines (WL) and vertical bitlines (BL) to form a flash memory ``block''\cite{Bez_2003}. A collection of such blocks makes up the flash memory chip. A schematic of the block structure of MLC flash memories is shown in Fig.~\ref{fig:flash_block}.

The two bits belonging to a MLC flash memory cell are separately mapped to logical units of programming, called pages. A page is also the smallest unit for program and read operations whereas a block is the smallest unit for the erase operation. The most significant bit (MSB) is mapped to the lower page while the least significant bit (LSB) is mapped to the upper page. The lower page bit of a cell always precedes the corresponding upper page bit in the programming order. We represent the four charge levels in MLC flash memory as 0, 1, 2, 3 in the increasing order of charge levels respectively. The corresponding \mbox{2-bit} patterns written to the lower (MSB) and upper (LSB) pages are `11', `10', `00' and `01' respectively as shown in~Fig.~\ref{fig:flash_block}.

%% file: tikz/flash_memory_block.tex
\tikzstyle{cell} = [rectangle, draw, minimum size=16pt,inner sep=0pt, outer sep=0pt]
\tikzstyle{cell_lp} = [rectangle, draw=blue, line width=0.5mm, minimum size=16pt,inner sep=0pt, outer sep=0pt]
\tikzstyle{cell_up} = [rectangle, draw=red, line width=0.5mm, minimum size=19pt,inner sep=0pt, outer sep=0pt]
\tikzstyle{legend_lp} = [rectangle, fill=blue, minimum size=8pt, inner sep=0pt, outer sep=0pt]
\tikzstyle{legend_up} = [rectangle, fill=red, minimum size=8pt, inner sep=0pt, outer sep=0pt]

\begin{tikzpicture}[yscale=0.3, xscale=1.0, node distance=0.1cm, auto, thick]

\node[cell] (cell-00) at (-5.0, -0.5) {3};
\node[cell] (cell-01) at (-5.0, -2.4) {2};
\node[cell] (cell-02) at (-5.0, -4.3) {1};
\node[cell] (cell-03) at (-5.0, -6.2) {0};

\node[cell] (cell-10) at (-4.0, -0.5) {\tb{0}\tr{1}};
\node[cell] (cell-11) at (-4.0, -2.4) {\tb{0}\tr{0}};
\node[cell] (cell-12) at (-4.0, -4.3) {\tb{1}\tr{0}};
\node[cell] (cell-13) at (-4.0, -6.2) {\tb{1}\tr{1}};

\node[align=left, below] at (-4.5, -7.5) {\footnotesize{Low Voltage}};
\node[align=left, above] at (-4.5,  0.5) {\footnotesize{High Voltage}};

\node[align=left, below] at (-4.5, -9.0) {\footnotesize{Cell Level to Bit Mapping}};

\node[legend_lp] (legend_lp) at (-5.5, 6.0) {};
\node[legend_up] (legend_up) at (-5.5, 4.0) {};
\node[align = left, right] at (-5.4, 5.8) {\footnotesize{Lower Page}};
\node[align = left, right] at (-5.4, 3.8) {\footnotesize{Upper Page}};

\node[cell] (cell-0-0) at (-1.0, 2.0) {\tb{1}\tr{0}};
\node[cell] (cell-0-1) at (0.0,  2.0) {\tb{1}\tr{1}};
\node[cell] (cell-0-2) at (1.0,  2.0) {\tb{0}\tr{1}};
\node[cell] (cell-0-3) at (2.0,  2.0) {\tb{1}\tr{0}};

\node[cell] (cell-1-0) at (-1.0, -1.0) {\tb{1}\tr{1}};
\node[cell] (cell-1-1) at (0.0, -1.0) {\tb{0}\tr{1}};
\node[cell] (cell-1-2) at (1.0, -1.0) {\tb{0}\tr{1}};
\node[cell] (cell-1-3) at (2.0, -1.0) {\tb{1}\tr{1}};

\node[cell] (cell-2-0) at (-1.0, -4.0) {\tb{0}\tr{1}};
\node[cell] (cell-2-1) at (0.0, -4.0) {\tb{1}\tr{0}};
\node[cell] (cell-2-2) at (1.0, -4.0) {\tb{0}\tr{0}};
\node[cell] (cell-2-3) at (2.0, -4.0) {\tb{1}\tr{1}};

\node[cell] (cell-3-0) at (-1.0, -7.0) {\tb{0}\tr{0}};
\node[cell] (cell-3-1) at (0.0, -7.0) {\tb{0}\tr{1}};
\node[cell] (cell-3-2) at (1.0, -7.0) {\tb{1}\tr{1}};
\node[cell] (cell-3-3) at (2.0, -7.0) {\tb{0}\tr{1}};

\draw[-] (cell-0-0) -- (cell-0-1);
\draw[-] (cell-0-1) -- (cell-0-2);
\draw[-] (cell-0-2) -- (cell-0-3);

\draw[-] (cell-1-0) -- (cell-1-1);
\draw[-] (cell-1-1) -- (cell-1-2);
\draw[-] (cell-1-2) -- (cell-1-3);

\draw[-] (cell-2-0) -- (cell-2-1);
\draw[-] (cell-2-1) -- (cell-2-2);
\draw[-] (cell-2-2) -- (cell-2-3);

\draw[-] (cell-3-0) -- (cell-3-1);
\draw[-] (cell-3-1) -- (cell-3-2);
\draw[-] (cell-3-2) -- (cell-3-3);

\draw[-] (cell-0-0) -- (cell-1-0);
\draw[-] (cell-1-0) -- (cell-2-0);
\draw[-] (cell-2-0) -- (cell-3-0);

\draw[-] (cell-0-1) -- (cell-1-1);
\draw[-] (cell-1-1) -- (cell-2-1);
\draw[-] (cell-2-1) -- (cell-3-1);

\draw[-] (cell-0-2) -- (cell-1-2);
\draw[-] (cell-1-2) -- (cell-2-2);
\draw[-] (cell-2-2) -- (cell-3-2);

\draw[-] (cell-0-3) -- (cell-1-3);
\draw[-] (cell-1-3) -- (cell-2-3);
\draw[-] (cell-2-3) -- (cell-3-3);


\draw[-, dashed] (-2.0, 2.0) -- (cell-0-0);
\draw[-, dashed] (-2.0, -1.0) -- (cell-1-0);
\draw[-, dashed] (-2.0, -4.0) -- (cell-2-0);
\draw[-, dashed] (-2.0, -7.0) -- (cell-3-0);

\draw[-, dashed] (2.8, 2.0) -- (cell-0-3);
\draw[-, dashed] (2.8, -1.0) -- (cell-1-3);
\draw[-, dashed] (2.8, -4.0) -- (cell-2-3);
\draw[-, dashed] (2.8, -7.0) -- (cell-3-3);

\draw[-, dashed] (-1.0, 5.0) -- (cell-0-0);
\draw[-, dashed] (0.0, 5.0) -- (cell-0-1);
\draw[-, dashed] (1.0, 5.0) -- (cell-0-2);
\draw[-, dashed] (2.0, 5.0) -- (cell-0-3);

\draw[-, dashed] (-1.0, -9.0) -- (cell-3-0);
\draw[-, dashed] (0.0, -9.0) -- (cell-3-1);
\draw[-, dashed] (1.0, -9.0) -- (cell-3-2);
\draw[-, dashed] (2.0, -9.0) -- (cell-3-3);

\node[draw=none] at (-2.1, 1.6) {$\textrm{WL}_{i-1}$};
\node[draw=none] at (-2.3, -1.4) {$\textrm{WL}_{i}$};
\node[draw=none] at (-2.1, -4.4) {$\textrm{WL}_{i+1}$};
\node[draw=none] at (-2.1, -7.4) {$\textrm{WL}_{i+2}$};

\node[draw=none] at (-0.7, 5.5) {$\textrm{BL}_{i-2}$};
\node[draw=none] at (0.3, 5.5) {$\textrm{BL}_{i-1}$};
\node[draw=none] at (1.1, 5.5) {$\textrm{BL}_{i}$};
\node[draw=none] at (2.3, 5.5) {$\textrm{BL}_{i+1}$};

\node[align=left, below] at (0.5, -9.0) {\footnotesize{MLC Flash Block Schematic}};




\end{tikzpicture}

%% file: sections/experiment_procedure.tex
\section{Experiment Procedure}
\label{sec:expt_proc}
To characterize and quantify the number and types of errors observed, we perform program/erase (P/E) cycling of the MLC flash memory chip under test which consists of repeated application of the following steps: 
\begin{enumerate}
\item Erase MLC flash memory blocks under test.
\item Program MLC flash memory pages (of blocks under test) with pseudo-random (PR) data generated using a  Mersenne-Twister pseudo-random number generator. The pseudo-random number generator is initialized with a randomly generated seed for every page in every P/E cycle.
\item Starting with the first cycle, perform a read operation on the MLC flash memory block(s) at intervals of every $100^{\textrm{th}}$ cycle. Record bit errors and their locations in the block. 
\end{enumerate}
We arbitrarily choose 4 contiguous blocks in an MLC flash memory chip for our experiments.
The MLC flash memory blocks are P/E cycled up to 10,000 P/E cycles and the experiments are performed at room temperature in a continuous manner with no extra wait time between the erase/program/read operations.

%% file: sections/error_characterization.tex
\section{Characterization of Errors in MLC Flash Memories}
\label{sec:cell_err_cap_ch}
The first step in the error characterization of a flash memory chip is to study its raw bit error rate (BER) performance when all the pages in all the blocks under test are programmed with pseudo-random data. This closely resembles the most common use in practice, where random data are stored and retrieved. Fig.~\ref{fig:rber} shows the average raw BER across the P/E cycles when all pages in each block are programmed for both the vendor-A and vendor-B flash memory chips. The raw BER is averaged over 4 blocks tested. Fig.~\ref{fig:rber} also shows the average raw BER separately for the lower and upper pages of the MLC flash memory. Although the lower page is expected to have a smaller BER compared to the upper page~\cite{Yaakobi_2012}, we observe that this is only the case up to a certain number of P/E cycles in the beginning and as the P/E cycle count increases, the lower page begins to show a larger number of errors than the upper page. This observation is consistent across both the vendor-A and vendor-B flash memory chips. Using empirical data from 20 blocks of the same flash memory chip, we have also observed consistent measured average raw BER estimates across all the P/E cycles.
\ifCLASSOPTIONonecolumn
	\begin{figure}
		\centering
		\includegraphics[width=0.7\textwidth]{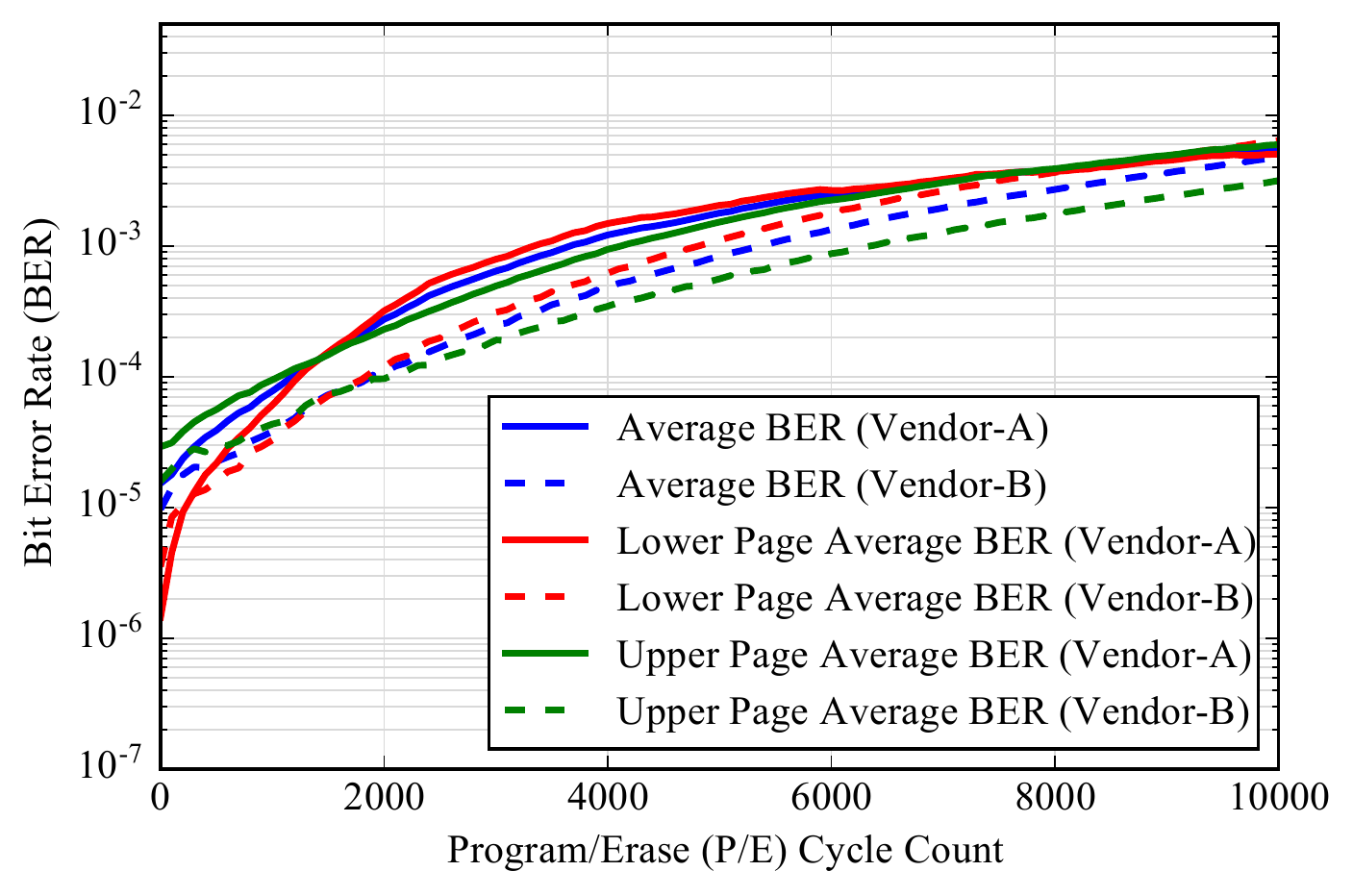}
		\caption{Measured average raw bit error rates over 4 blocks of \mbox{vendor-A} and \mbox{vendor-B} chips.}
		\label{fig:rber}
	\end{figure}
\else
	\begin{figure}
		\centering
		\includegraphics[width=0.47\textwidth]{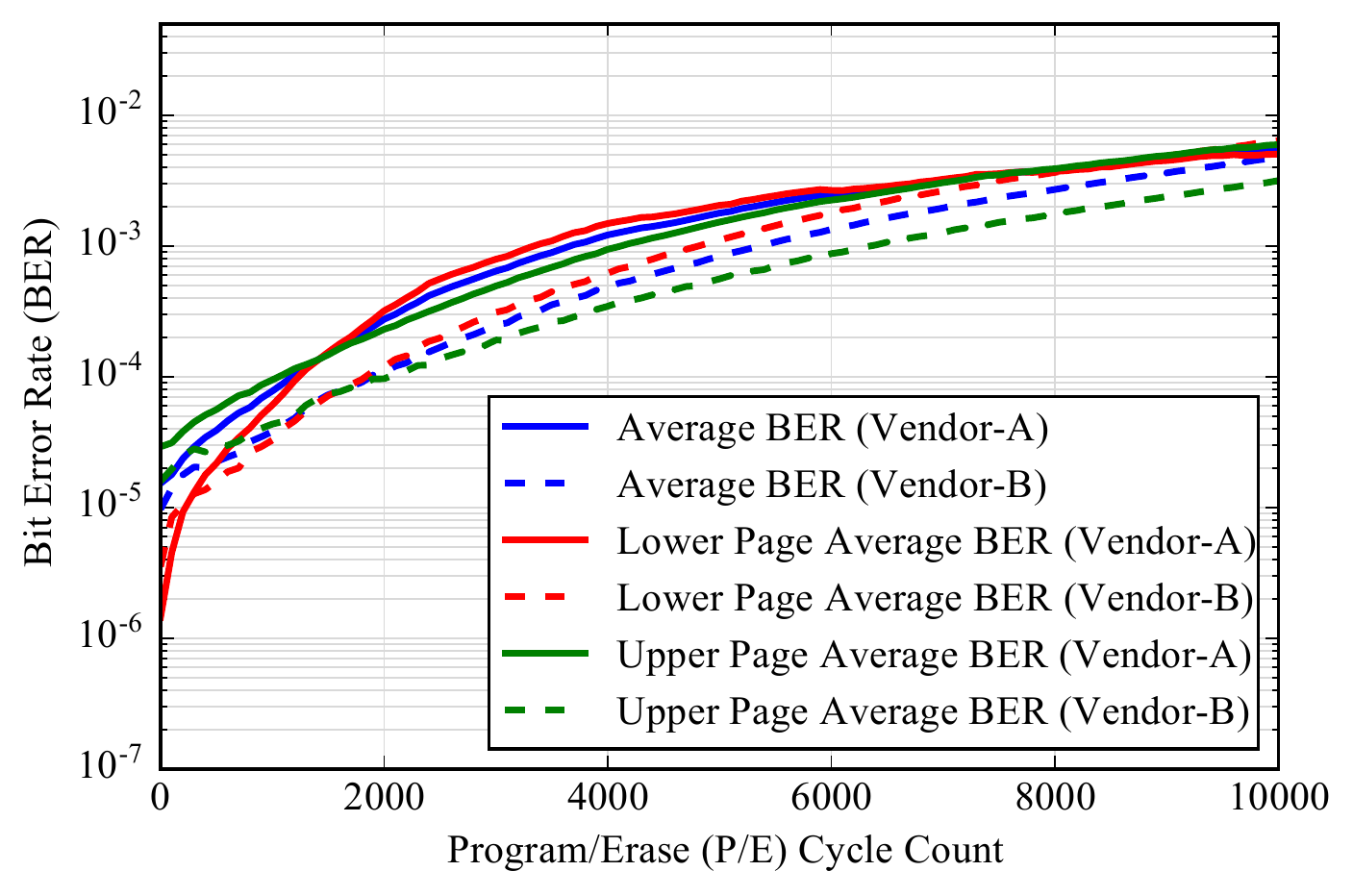}
		\caption{Measured average raw bit error rates over 4 blocks of \mbox{vendor-A} and \mbox{vendor-B} chips.}
		\label{fig:rber}
	\end{figure}
\fi
\ifCLASSOPTIONonecolumn
	\begin{figure}
		\centering
		\includegraphics[width=0.7\textwidth]{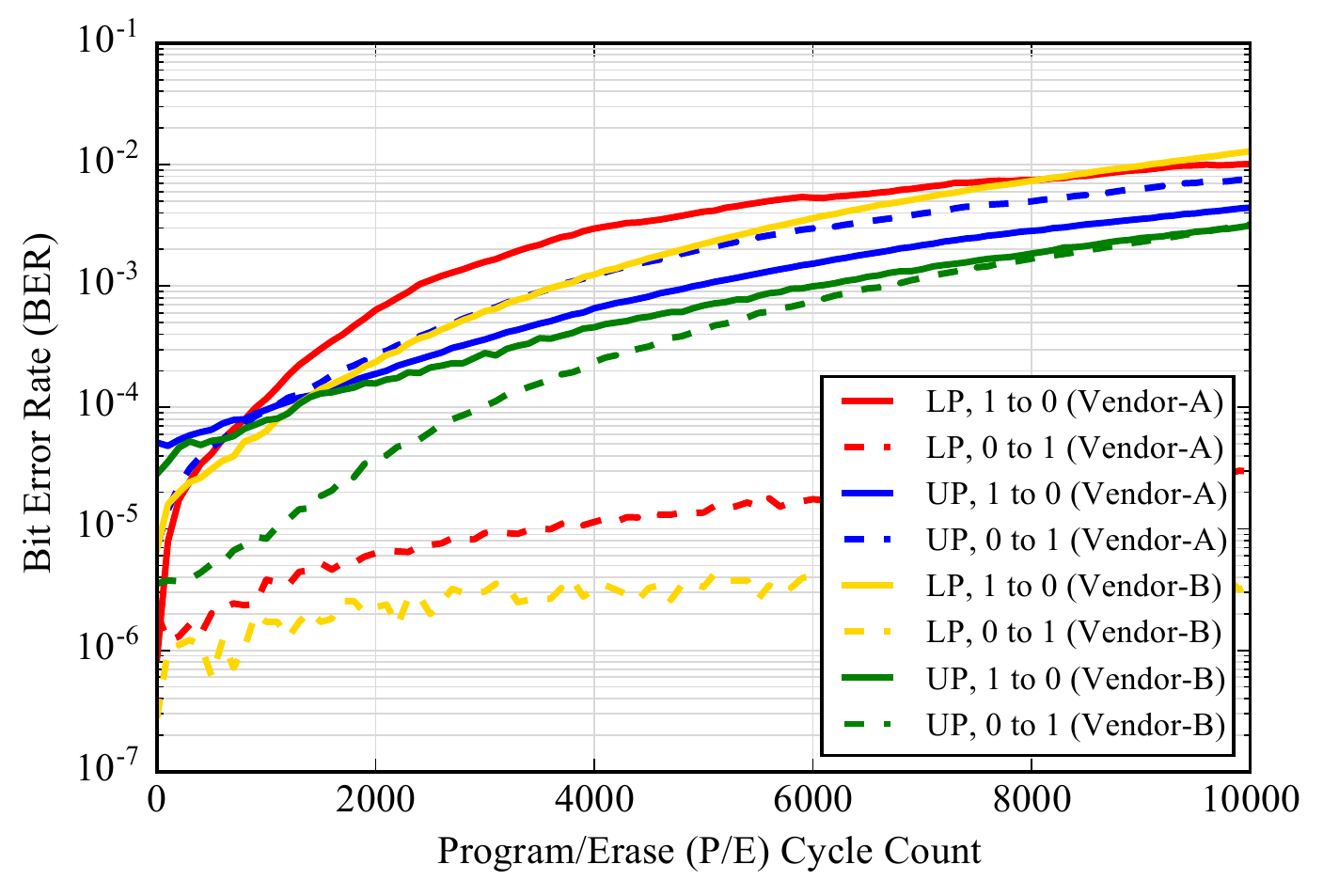}
		\caption{Average raw bit error rates corresponding to specific bit errors
		in the lower pages (LP) and upper pages (UP) over 4 blocks of \mbox{vendor-A} and \mbox{vendor-B} chips.}
		\label{fig:asym_flash_channel}
	\end{figure}
\else
	\begin{figure}
		\centering
		\includegraphics[width=0.47\textwidth]{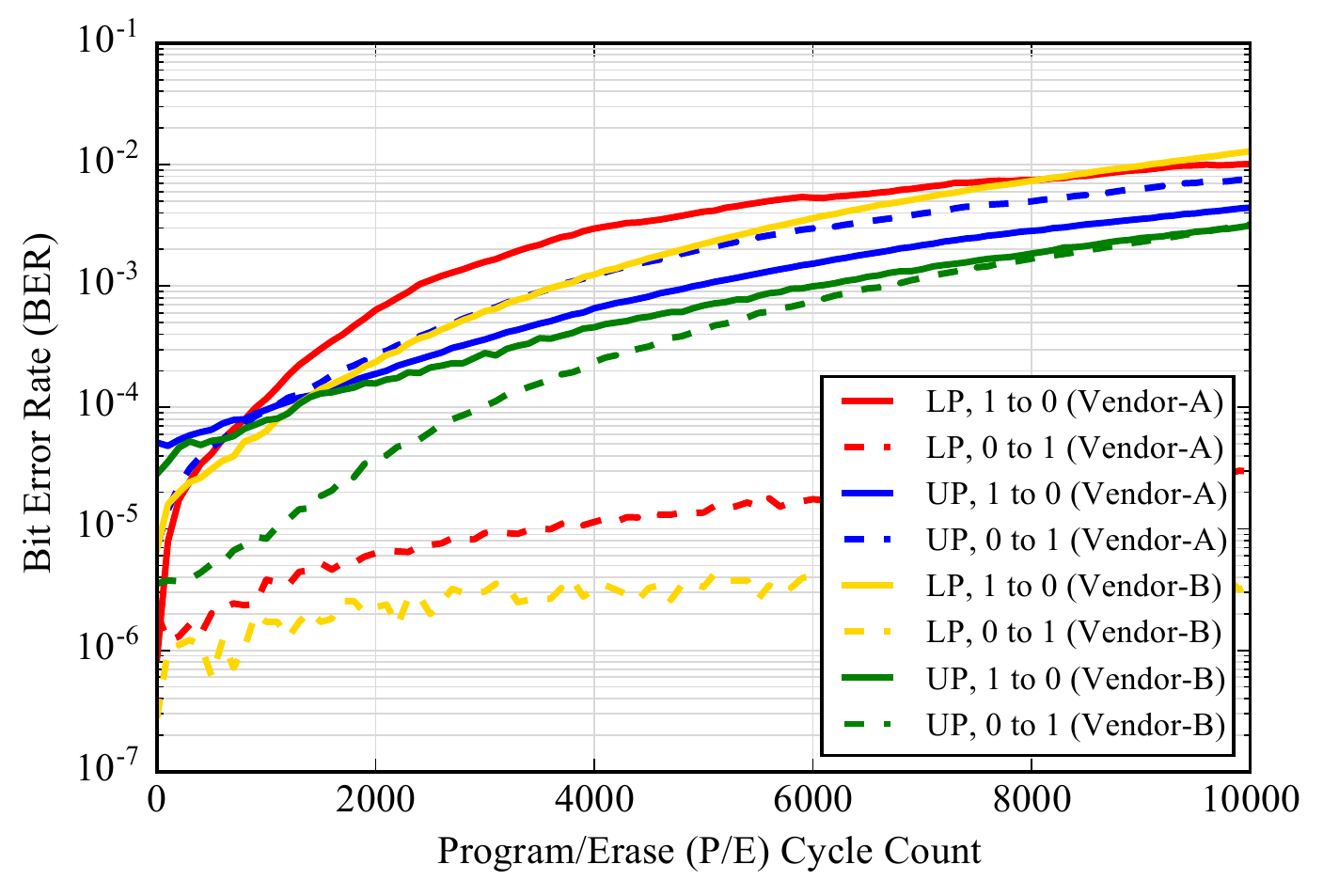}
		\caption{Average raw bit error rates corresponding to specific bit errors
		in the lower pages (LP) and upper pages (UP) over 4 blocks of \mbox{vendor-A} and \mbox{vendor-B} chips.}
		\label{fig:asym_flash_channel}
	\end{figure}
\fi

\begin{table}[!ht]
\caption{Frequency of cell (symbol) errors measured as a percentage of total number of cell errors observed across all P/E cycles when all 4 blocks are programmed with pseudo-random data.}
\label{table:percent_sym_errors}
\centering
\bgroup

\ifCLASSOPTIONonecolumn
	\def\arraystretch{1.0}
	\tabcolsep=0.11cm
	\resizebox{0.3\columnwidth}{!}{
		\begin{tabular}{|c|c c c c|}
			\hline
			\multicolumn{5}{|c|}{\textbf{Vendor-A}} \\
			\hline
			\textbf{Write Cell} & \multicolumn{4}{c|}{\textbf{Read Cell Values}} \\
			\hhline{~----}
			\textbf{Values} & \textbf{11} & \textbf{10} & \textbf{00} & \textbf{01} \\
			\hline
			\textbf{11} & 0.00 & 17.25 & 0.08 & 2.57 \\
			\hline
			\textbf{10} & 0.19 & 0.00 & 48.19 & 0.74 \\
			\hline
			\textbf{00} & 0.00 & 0.14 & 0.00 & 30.61 \\
			\hline
			\textbf{01} & 0.00 & 0.03 & 0.20 & 0.00 \\\hline
		\end{tabular}
	}
	\resizebox{0.3\columnwidth}{!}{
		\begin{tabular}{|c|c c c c|}
			\hline
			\multicolumn{5}{|c|}{\textbf{Vendor-B}} \\
			\hline
			\textbf{Write Cell} & \multicolumn{4}{c|}{\textbf{Read Cell Values}} \\
			\hhline{~----}
			\textbf{Values} & \textbf{11} & \textbf{10} & \textbf{00} & \textbf{01} \\
			\hline
			\textbf{11} & 0.00 & 18.39 & 0.03 & 4.01 \\
			\hline
			\textbf{10} & 0.07 & 0.00 & 62.22 & 1.84 \\
			\hline
			\textbf{00} & 0.00 & 0.06 & 0.00 & 13.39 \\
			\hline
			\textbf{01} & 0.00 & 0.00 & 0.00 & 0.00 \\
			\hline
		\end{tabular}
	}
\else
	\def\arraystretch{1.5}
	\tabcolsep=0.11cm
	\resizebox{0.49\columnwidth}{!}{
		\begin{tabular}{|c|c c c c|}
			\hline
			\multicolumn{5}{|c|}{\textbf{Vendor-A}} \\
			\hline
			\textbf{Write Cell} & \multicolumn{4}{c|}{\textbf{Read Cell Values}} \\
			\hhline{~----}
			\textbf{Values} & \textbf{11} & \textbf{10} & \textbf{00} & \textbf{01} \\
			\hline
			\textbf{11} & 0.00 & 17.25 & 0.08 & 2.57 \\
			\hline
			\textbf{10} & 0.19 & 0.00 & 48.19 & 0.74 \\
			\hline
			\textbf{00} & 0.00 & 0.14 & 0.00 & 30.61 \\
			\hline
			\textbf{01} & 0.00 & 0.03 & 0.20 & 0.00 \\\hline
		\end{tabular}
	}
	\resizebox{0.49\columnwidth}{!}{
		\begin{tabular}{|c|c c c c|}
			\hline
			\multicolumn{5}{|c|}{\textbf{Vendor-B}} \\
			\hline
			\textbf{Write Cell} & \multicolumn{4}{c|}{\textbf{Read Cell Values}} \\
			\hhline{~----}
			\textbf{Values} & \textbf{11} & \textbf{10} & \textbf{00} & \textbf{01} \\
			\hline
			\textbf{11} & 0.00 & 18.39 & 0.03 & 4.01 \\
			\hline
			\textbf{10} & 0.07 & 0.00 & 62.22 & 1.84 \\
			\hline
			\textbf{00} & 0.00 & 0.06 & 0.00 & 13.39 \\
			\hline
			\textbf{01} & 0.00 & 0.00 & 0.00 & 0.00 \\
			\hline
		\end{tabular}
	}
\fi
\egroup
\end{table}


We also record the specific cell (symbol) errors corresponding to all the bit errors observed. Table~\ref{table:percent_sym_errors} shows the frequencies of all possible cell errors as a percentage of the total number of cell errors observed across all the blocks in all the P/E cycles. The corresponding average cell error probabilities across all P/E cycles are $\ca 4.16 \times 10^{-3}$ and $\ca 2.71 \times 10^{-3}$ for vendor-A and vendor-B chips respectively. We observe that the level 1 to 2 cell error ``10~(1)~$\rightarrow$~00~(2)'' is the most dominant for both vendor-A and vendor-B chips. This observation explains why the lower page average raw BER is worse than the upper page average raw BER as shown earlier in Fig.~\ref{fig:rber}. We also note that the three adjacent level cell errors ``10~(1)~$\rightarrow$~00~(2)'', ``11~(0)~$\rightarrow$~10~(1)'' and ``00~(2)~$\rightarrow$~01~(3)'' are the most frequent and together make up about 96\% and 94\% of all the cell errors observed for the vendor-A and vendor-B chips respectively.
Such knowledge about dominant cell errors can be very useful in utilizing ECC redundancy more effectively. This was demonstrated in~\cite{Yaakobi_2010}, where the authors designed two BCH codes with different error correction capabilities for the lower and upper pages of an MLC flash memory and proposed a stagewise combined decoding algorithm for both pages. Their scheme gave better results than using a single BCH code independently for all pages.
\subsection{Asymmetry of Bit Errors in MLC Flash Memories}
\label{subsec:asymm_bit_errors}
Fig.~\ref{fig:asym_flash_channel} shows the asymmetry of bit errors in MLC flash memories. We present the average raw BERs corresponding to the specific types of bit errors i.e., $\ZOerror$ and $\OZerror$ bit errors, in the lower and upper pages of both vendor-A and vendor-B MLC flash memory chips. While there is a  high degree of asymmetry in the lower page bit errors throughout the P/E cycle range, the degree of asymmetry in the upper page bit errors is much lower. This agrees well with the observations in~Table~\ref{table:percent_sym_errors}, where the dominant cell errors imply a large proportion of $\OZerror$ bit errors in the lower page and comparable proportions of $\ZOerror$ and $\OZerror$ bit errors in the upper page. This asymmetry in bit errors in both the lower and upper pages also reflects the dominance of data dependent inter-cell interference (ICI) errors i.e., the middle cells in the cell level data patterns $303$, $313$ and $323$ across wordlines are highly susceptible to errors~\cite{Taranalli_2015}.
\subsection{Characterization of Number of Bit Errors per Frame}
As we want to develop parametric channel models for MLC flash memories which provide an accurate representation of the empirically observed bit errors and enable accurate ECC FER performance estimation, we study the distribution of the number of bit errors per frame parameter. This is the key factor in determining the FER performance of an ECC with a specified error correction capability of $t$ number of bit errors per frame.

From the error data collected during P/E cycling experiments, we obtain the sample counts of the number of bit errors per frame for $\ZOerror$ and $\OZerror$ bit errors in both the lower and upper pages by choosing a fixed frame length of $N = 8192$ bits. This choice of the frame length is representative of the large ECC frame lengths used in practice, while still being small enough to ensure sufficient empirical data can be collected easily. Commonly used ECC frame lengths range from $8192$ to $32768$ bits and multiple ECC frames are written to a single flash memory page in practice.
The sample mean and variance statistics of the number of bit errors per frame are computed using the sample counts and are shown in~Table~\ref{table:sample_mean_var_table} for both vendor-A and vendor-B chips. We also plot two dimensional (2D) maps showing the number of bit errors for every frame in a single block of MLC flash memory at 8,000 P/E cycles~in~Fig.~\ref{fig:error_maps_8000_vendor_ab}. The 2D maps are obtained by stacking horizontally, the bit error counts in frames belonging to a page, and then stacking vertically all the pages belonging to a single block.
From Table~\ref{table:sample_mean_var_table} and Fig.~\ref{fig:error_maps_8000_vendor_ab}, we clearly observe that the variance in the number of bit errors per frame is much larger than the mean i.e., the experiment data is overdispersed with respect to a binomial distribution, $\textrm{Binomial}(n, p)$,  typically used to model count data whose mean and variance are approximately equal when $p$ is small.
%
\begin{table}[!ht]
	\caption{Sample mean and variance of the number of bit errors per frame obtained from empirical data for lower and upper pages across P/E cycles when all 4 blocks are programmed with pseudo-random data. Frame length $N = 8192$.}
	\label{table:sample_mean_var_table}
	\centering
	\bgroup
	\ifCLASSOPTIONonecolumn
		\begin{tabular}{|c|c c |c c|c c|c c|}
			\hline
			\textbf{P/E} & \multicolumn{4}{|c|}{\textbf{Vendor-A}} & \multicolumn{4}{|c|}{\textbf{Vendor-B}}\\
			\hhline{~--------}
			\textbf{Cycles} & \multicolumn{2}{c|}{\textbf{Lower Page}} & \multicolumn{2}{c|}{\textbf{Upper Page}} & \multicolumn{2}{c|}{\textbf{Lower Page}} & \multicolumn{2}{c|}{\textbf{Upper Page}} \\
			\hhline{~--------}
			& \textbf{Mean} & \textbf{Variance} & \textbf{Mean} & \textbf{Variance} & \textbf{Mean} & \textbf{Variance} & \textbf{Mean} & \textbf{Variance} \\
			\hline
			2000 & 2.63 & 3.08 & 1.90 & 2.17 & 0.98 & 1.05 & 0.79 & 0.86 \\
			\hline
			4000 & 12.21 & 18.70 & 7.76 & 9.84 & 5.10 & 6.97 & 2.84 & 3.66 \\
			\hline
			6000 & 21.90 & 46.71 & 18.43 & 30.06 & 14.85 & 29.64 & 7.18 & 10.23 \\
			\hline
			8000 & 30.55 & 75.89 & 32.01 & 66.43 & 30.03 & 84.81 & 14.46 & 24.37 \\
			\hline
			10000 & 41.37 & 111.35 & 48.88 & 125.99 & 52.61 & 216.95 & 26.06 & 51.30 \\
			\hline
		\end{tabular}
	\else
		\def\arraystretch{1.5}
		\tabcolsep=0.11cm
		\resizebox{0.98\columnwidth}{!}{
		\begin{tabular}{|c|c c |c c|c c|c c|}
			\hline
			\textbf{P/E} & \multicolumn{4}{|c|}{\textbf{Vendor-A}} & \multicolumn{4}{|c|}{\textbf{Vendor-B}}\\
			\hhline{~--------}
			\textbf{Cycles} & \multicolumn{2}{c|}{\textbf{Lower Page}} & \multicolumn{2}{c|}{\textbf{Upper Page}} & \multicolumn{2}{c|}{\textbf{Lower Page}} & \multicolumn{2}{c|}{\textbf{Upper Page}} \\
			\hhline{~--------}
			& \textbf{Mean} & \textbf{Variance} & \textbf{Mean} & \textbf{Variance} & \textbf{Mean} & \textbf{Variance} & \textbf{Mean} & \textbf{Variance} \\
			\hline
			2000 & 2.63 & 3.08 & 1.90 & 2.17 & 0.98 & 1.05 & 0.79 & 0.86 \\
			\hline
			4000 & 12.21 & 18.70 & 7.76 & 9.84 & 5.10 & 6.97 & 2.84 & 3.66 \\
			\hline
			6000 & 21.90 & 46.71 & 18.43 & 30.06 & 14.85 & 29.64 & 7.18 & 10.23 \\
			\hline
			8000 & 30.55 & 75.89 & 32.01 & 66.43 & 30.03 & 84.81 & 14.46 & 24.37 \\
			\hline
			10000 & 41.37 & 111.35 & 48.88 & 125.99 & 52.61 & 216.95 & 26.06 & 51.30 \\
			\hline
		\end{tabular}
		}
	\fi
	\egroup
\end{table}

\ifCLASSOPTIONonecolumn
	\begin{figure}[!ht]
		\centering
		\includegraphics[width=0.7\textwidth]{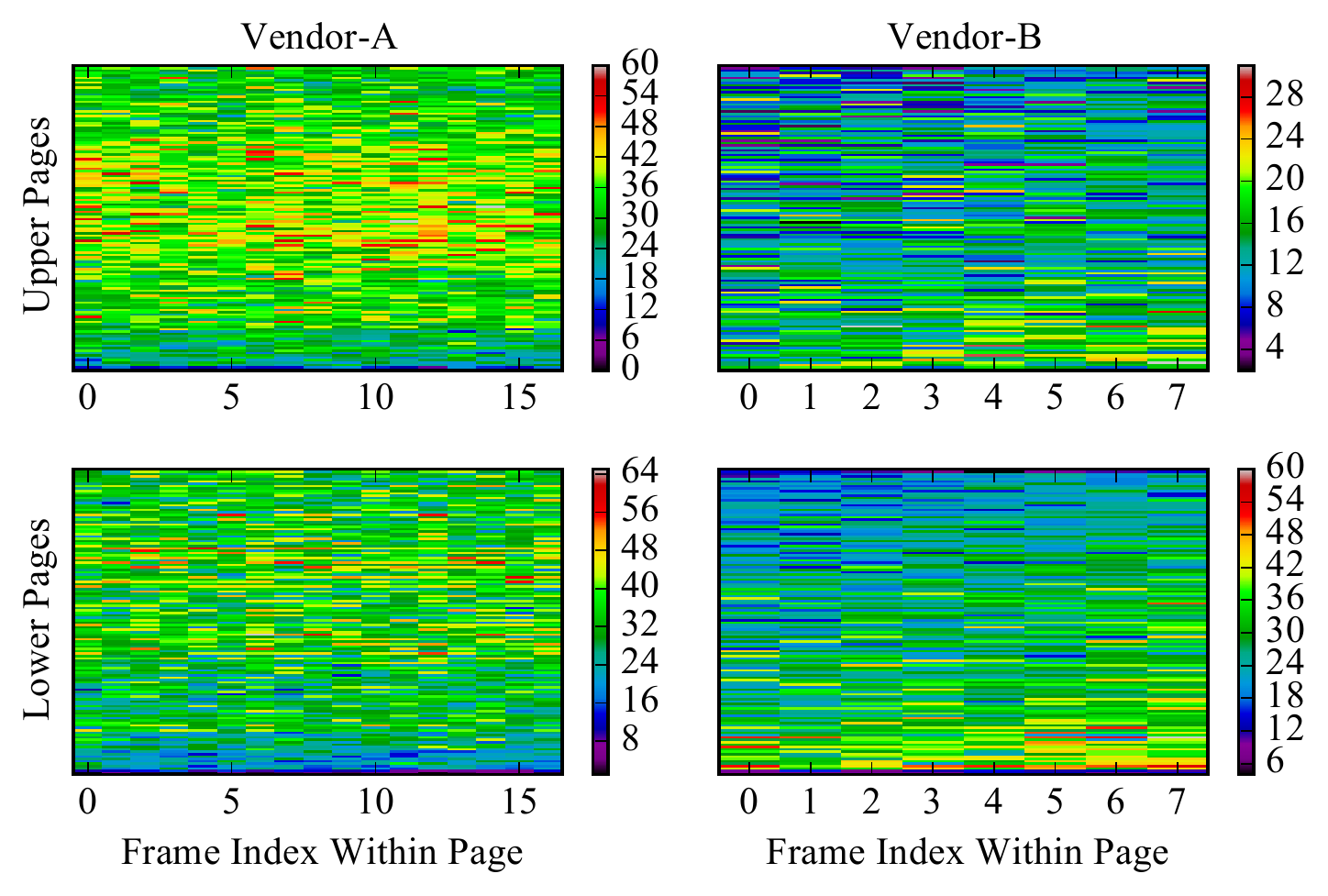}
		\caption{Two dimensional maps of bit error counts in frames of lower and upper pages in a single block of MLC flash memory chips from \mbox{vendor-A} and \mbox{vendor-B} at 8,000 P/E cycles. }
		\label{fig:error_maps_8000_vendor_ab}
	\end{figure}
\else
	\begin{figure}[!ht]
		\centering
		\includegraphics[width=0.47\textwidth]{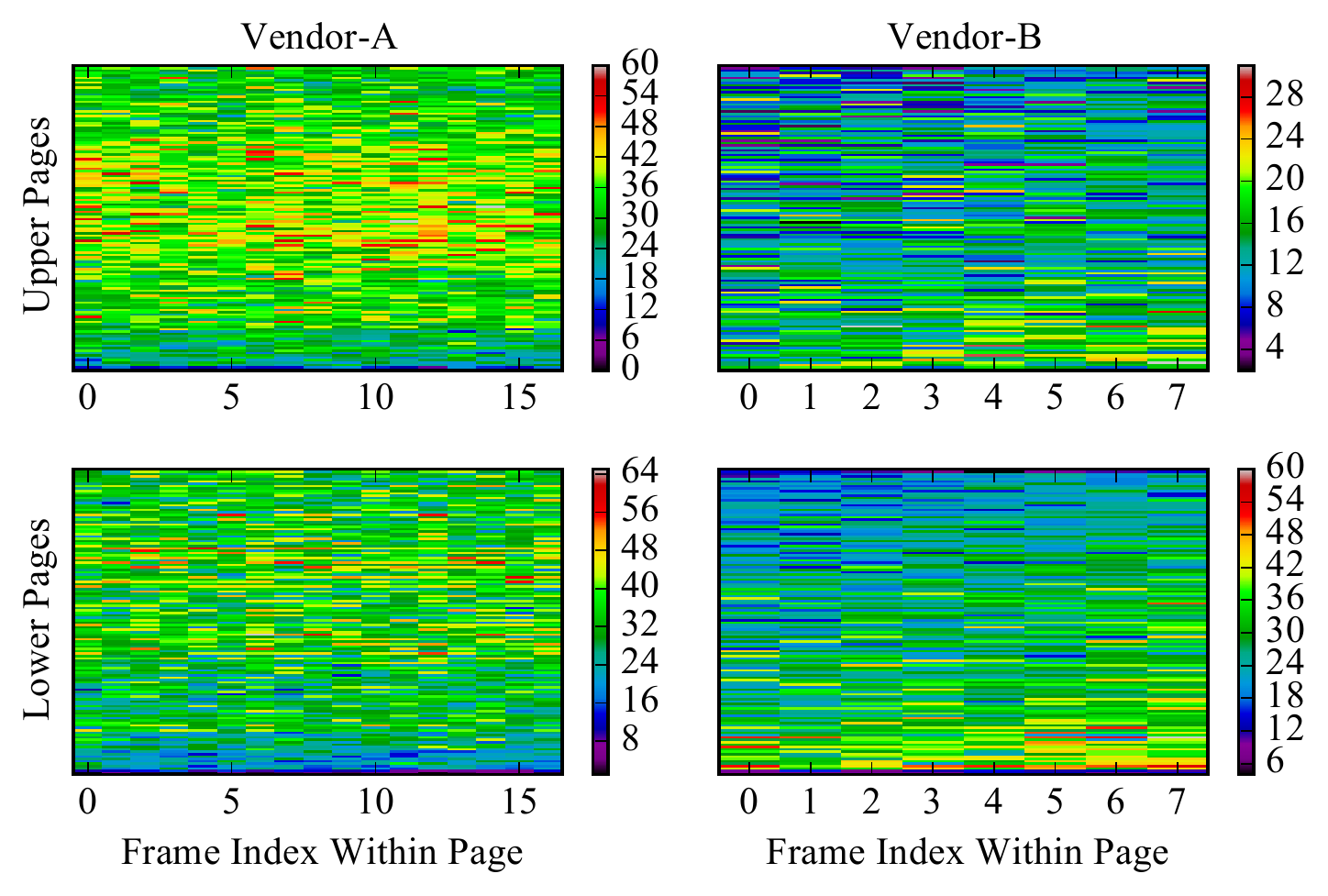}
		\caption{Two dimensional maps of bit error counts in frames of lower and upper pages in a single block of MLC flash memory chips from \mbox{vendor-A} and \mbox{vendor-B} at 8,000 P/E cycles. }
		\label{fig:error_maps_8000_vendor_ab}
	\end{figure}
\fi

%% file: sections/dmc_for_flash.tex
\section{Channel Models for MLC Flash Memories}
\label{sec: dmc_for_mlc_flash}
In this section, first we study the suitability of well known discrete memoryless channel (DMC) models such as the \mbox{4-ary} DMC, the BSC and the BAC, to represent the bit errors observed in the MLC flash memory channel. Among the DMC models, a per page BAC (\mbox{2-BAC}) model appears to align well with our empirical error characterization results. However we show through analysis as well as empirical results that the per page BAC model is unable to fit the empirical distribution of the number of bit errors per frame and is not a good model for ECC FER performance estimation. This is due to the interdependence of mean and variance statistics of the number of bit errors per frame for a BAC where the number of $\ZOerror$ and $\OZerror$ errors are modeled as binomial distributions. The binomial distribution is a single parameter (degree of freedom) distribution, hence its mean and variance cannot be chosen independently. Thus the binomial distribution is unable to accurately model the overdispersed empirical error data as described in the previous section. A natural next choice is to consider the normal approximation to the binomial distribution which provides two parameters (degrees of freedom) for modeling the observed mean and variance statistics independently. However we observe that the normal approximation based channel model does not accurately fit the shape of the empirical data distribution. Another commonly used probability distribution to model overdispersed data with respect to a binomial distribution is the beta-binomial distribution~\cite{Skellam_1948, Lindsey_1998}. Hence we propose a discrete channel model based on the beta-binomial distribution for the lower and upper pages referred to as the \mbox{2-BBM} channel model. We show that this model fits the empirical distribution of the number of bit errors per frame and provides accurate ECC FER performance estimation. We also present simple approximations of the 2-BAC model based on the normal and Poisson probability distributions. Although these approximations are able to fit the empirical distribution of the number of bit errors per frame better than the 2-BAC model, they are not as good a fit as the proposed 2-BBM channel model.
\subsection{Definitions and Notation}
Let $K$ represent the total number of bit errors in a frame of length $N$ bits. Let $K_m$ be the total number of bit errors in a frame of $N$ bits which consists of $m$ zeros and $N-m$ ones. The relationship between probability distributions of $K$ and $K_m$ is given by
\begin{IEEEeqnarray}{L}
    \Pr(K = k) = \sum_{m=0}^{N}\frac{{N \choose m}}{2^{N}}~\Pr(K_m = k)
    \label{eqn:k_dist}
\end{IEEEeqnarray}
where $\frac{{N \choose m}}{2^{N}}$ represents the probability of observing exactly $m$ zeros in a frame of $N$ bits.
$K_m$ can be represented as the sum of the number of $\ZOerror$ and $\OZerror$ bit errors as
\begin{IEEEeqnarray}{L}
    K_m = K_m^{(0)} + K_{N-m}^{(1)}
    \label{eqn:km_sum}
\end{IEEEeqnarray}
where $\kcount{m}{0}$ and $\kcount{N-m}{1}$ denote the number of $\ZOerror$ and $\OZerror$ bit errors respectively.
$K$ can also be represented as the sum of the total number of $\ZOerror$ and $\OZerror$ bit errors as
\begin{IEEEeqnarray}{rCl}
	K & = & K^{(0)} + K^{(1)}~~~\textrm{where,} \\
	\Pr(K^{(u)} = k) & = & \sum_{m=k}^{N}\frac{{N \choose l}}{2^{N}}~\Pr(\kcount{l}{u} = k) \label{eqn:k0_dist}
\end{IEEEeqnarray}
Note that $u \in \{ 0, 1\}$ where $l = m + (N - 2m)u$.
We use $\E[X]$ and $\Var[X]$ to denote the expected value (mean) and the variance of a random variable $X$ respectively. We use $X~|~Y$ to denote ``$X$ given $Y$''.
\subsection{Candidate Discrete Memoryless Channel (DMC) Models}
The primary error mechanism in MLC flash memories is at the cell level and hence the \mbox{4-ary} DMC model with 4 inputs and 4 outputs
can naturally account for all the cell level errors. This \mbox{4-ary} DMC model requires 16 parameters (only 12 independent parameters) which are the cell level transition probabilities
and these parameters can be easily estimated from experiment data such as that shown in Table~\ref{table:percent_sym_errors}. However the \mbox{4-ary} DMC model is not useful in practice as the logical unit of progam/read operations in current MLC flash memory applications is a binary page. Hence any practically applicable channel model would have to treat the errors in the lower and upper pages of the MLC flash memory independently, even though it is clear that the errors occur at the cell level and hence the lower and upper page bit errors are not independent.
%

A simpler more commonly used DMC model is the 2-BSC model where two independent BSCs are used to represent the bit errors occuring in the lower and upper pages. The advantage of using the BSC model for each page independently is that it is simple and well studied, with a variety of error correction coding (ECC) techniques available for transmission over the BSC. However, based on our error characterization results in Section~\ref{sec:cell_err_cap_ch}, the bit errors in MLC flash memories during P/E cycling are mostly asymmetric in nature. Therefore, the BSC is clearly not an accurate model to represent the bit errors in MLC flash memories. A numerical comparison of estimated capacities of the 4-ary DMC model and the 2-BSC model was presented in~\cite{Taranalli_2015}, where it was observed that the 4-ary DMC model provides a significant capacity gain compared to the 2-BSC model for MLC flash memories.
\subsection{The 2-Binary Asymmetric Channel (2-BAC) Model}
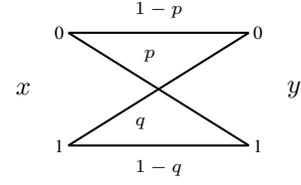
\begin{figure}
    \centering
    \begin{tikzpicture}[yscale=0.3, xscale=1.2, node distance=0.2cm, auto, thick]
        \draw[-, thick] (-1.0, 0.0) -- (1.0, 0.0);
        \draw[-, thick] (-1.0, -5.0) -- (1.0, -5.0);
        \draw[-, thick] (-1.0, 0.0) -- (1.0, -5.0);
        \draw[-, thick] (-1.0, -5.0) -- (1.0, 0.0);

        \node at (0.0, 1.0) {\scriptsize{$1 - p$}};
        \node at (0.0, -6.0) {\scriptsize{$1 - q$}};
        \node at (-0.1, -1.0) {\scriptsize{$p$}};
        \node at (-0.2, -4.0) {\scriptsize{$q$}};

        \node at (-1.1, 0.0) {\scriptsize{0}};
        \node at (-1.1, -5.0) {\scriptsize{1}};
        \node at (1.1, 0.0) {\scriptsize{0}};
        \node at (1.1, -5.0) {\scriptsize{1}};

        \node at (-1.5, -2.5) {$x$};
        \node at (1.5, -2.5) {$y$};

    \end{tikzpicture}
    \vspace*{-0.9em}
    \caption{Binary asymmetric channel}
    \label{fig:bac}
\end{figure}

Based on the asymmetry of bit errors observed in MLC flash memories (Section~\ref{sec:cell_err_cap_ch}), we propose a per page BAC model called the 2-BAC model where two independent BAC models are used to represent the bit errors occuring in the lower and upper pages. 
The 2-BAC model is a parametric model with 4 parameters which are the probabilities of $\ZOerror$ and $\OZerror$ errors in lower and upper page BACs, $p_0^{(l)}$, $p_1^{(l)}$ and $p_0^{(u)}$, $p_1^{(u)}$.
%
%
For a theoretical evaluation, we mainly compare the mean and variance statistics of the number of bit errors per frame corresponding to a BAC model with the empirically observed sample mean and variances shown in Table~\ref{table:sample_mean_var_table}.
We consider a BAC as shown in Fig.~\ref{fig:bac}, where $p$ is the probability of $\ZOerror$ error and $q$ is the probability of $\OZerror$ error. Next, we derive closed form expressions for the mean, $\E[K]$, and the variance, $\Var[K]$, of the number of bit errors per frame corresponding to a BAC model.
For the BAC model, $\kcount{m}{0}$ and $\kcount{N-m}{1}$ are distributed according to the binomial probability distribution and are independent i.e.,
\ifCLASSOPTIONonecolumn
    \begin{IEEEeqnarray}{rCl}
       \kcount{m}{0} &\sim& \textrm{Binomial}(m, p) \label{eqn:x_binom} \\
       \kcount{N-m}{1} &\sim& \textrm{Binomial}(N-m, q) \label{eqn:y_binom} \\
       \kcount{m}{0} &\indep& \kcount{N-m}{1} \label{eqn:xy_indep}
   \end{IEEEeqnarray}
\else
    \begin{IEEEeqnarray}{rCl}
       \kcount{m}{0} &\sim& \textrm{Binomial}(m, p) \label{eqn:x_binom} \\
       \kcount{N-m}{1} &\sim& \textrm{Binomial}(N-m, q) \label{eqn:y_binom} \\
       \kcount{m}{0} &\indep& \kcount{N-m}{1} \label{eqn:xy_indep}
   \end{IEEEeqnarray}
\fi
The mean and the variance of $\kcount{m}{0}$ are given by
\begin{IEEEeqnarray}{rCl}
    \E[\kcount{m}{0}] &=& mp \\
    \Var[\kcount{m}{0}] &=& mp(1-p)
\end{IEEEeqnarray}
and those of $\kcount{N-m}{1}$ are given by
\begin{IEEEeqnarray}{rCl}
    \E[\kcount{N-m}{1}] &=& (N-m)q \\
    \Var[\kcount{N-m}{1}] &=& (N-m)q(1-q).
\end{IEEEeqnarray}

\begin{IEEEprop}
	The mean and the variance of $K$ for a BAC model are given by
	\begin{IEEEeqnarray}{rCl}
		\E[K] & = & \frac{N}{2}(p + q) \\
		\Var[K] & = & \frac{N}{2}\Big((p + q) - pq - \frac{1}{2}(p^2 + q^2) \Big).
	\end{IEEEeqnarray}
    \label{prop:bac_k_mean_var}
\end{IEEEprop}
\begin{IEEEproof}
	See Appendix~\ref{app:proof_bac_k_mean_var}.
\end{IEEEproof}
%
The parameters of the BAC model $p$ and $q$ are estimated as the average $\ZOerror$ and $\OZerror$ bit error rates obtained from experimental data corresponding to a particular P/E cycle point in the flash memory lifetime. An algorithmic description of the BAC model is presented in~Algorithm~\ref{algo:bac_channel_model}.
\begin{algorithm}[!ht]
	\caption{\mbox{BAC} Model Implementation}
	\label{algo:bac_channel_model}
	\algsetup{
		linenosize=\small
	}
	\begin{algorithmic}[1]
        \REQUIRE Input frame $\mathbf{x}$ of length $N$, BAC model parameters $(p, q)$.
        \ENSURE Data frame with errors $\mathbf{y}$.
        \FOR{$x_i \in \mathbf{x}$}
            \STATE Generate random sample $u~\sim~\textrm{Uniform}[0, 1]$.
            \STATE \textbf{if} $x_i = 0$ \textbf{then} $t = p$ \textbf{else} $t = q$.
            \STATE \textbf{if} $u \leq t$ \textbf{then} $e_i = 1$ \textbf{else} $e_i = 0$.
            \STATE $y_i = x_i \oplus e_i$.
        \ENDFOR
    \end{algorithmic}
\end{algorithm}

Using the results of Proposition~\ref{prop:bac_k_mean_var}, we compute $\E[K]$ and $\Var[K]$ for a BAC model as follows.  For example, at 8,000 P/E cycles for the upper page BAC model for vendor-A chip, we have $p = 4.97 \times 10^{-3}$ and $q = 2.84 \times 10^{-3}$ and assuming $N = 8192$, we get $\E[K] = 32.01$ and $\Var[K] = 32.02$\@.
Comparing $\E[K]$ and $\Var[K]$ to the sample mean and variance of $K$ recorded using experimental data as shown in Table~\ref{table:sample_mean_var_table}, we observe that the BAC model is unable to account for the large observed sample variance. For small values of $p$ and $q$, from Proposition~\ref{prop:bac_k_mean_var}, we have $\Var[K] \approx \E[K]$.
Therefore, the BAC model is not a good fit for the observed empirical probability distribution of $K$ as shown in Fig.~\ref{fig:cdfs_8000_vendor_a} and Fig.~\ref{fig:cdfs_8000_vendor_b} for \mbox{vendor-A} and \mbox{vendor-B} flash memory chips, respectively.
As the $\Var[K]$ is much less than the observed sample variance, the 2-BAC model for MLC flash memory is expected to provide a more optimistic estimate of the ECC FER performance when compared to the actual performance. We discuss this in more detail in~Section~\ref{sec:empirical_results}. However, note that the 2-BAC model does provide an accurate estimate of the average raw BER which is given by $\frac{\E[K]}{N}$. This shows that the ability to accurately estimate/predict the average raw BER is not the sole criterion for a good MLC flash memory channel model.
\subsection{The 2-Beta-Binomial (2-BBM) Channel Model}
As mentioned in Section~\ref{sec:cell_err_cap_ch}, the empirically observed sample mean and variance estimates show that the number of bit errors per frame data is overdispersed with respect to the binomial distribution. This is the major reason for the poor fit of the 2-BAC model discussed in the previous subsection. To account for the overdispersion, we propose a channel model for MLC flash memories based on the beta-binomial probability distribution called the 2-Beta-Binomial (2-BBM) channel model.

The beta-binomial probability distribution was first proposed in~\cite{Skellam_1948} as the probability distribution for counts resulting from a binomial distribution if the probability of success varies according to the beta distribution between sets of trials. Using empirical data, it was also shown in~\cite{Skellam_1948} that the beta-binomial probability distribution is a good fit for overdispersed binomial data. Lindsey~et~al\@.~\cite{Lindsey_1998} studied the beta-binomial probability distribution based model in fitting overdispersed human sex ratio in families data and it was found to be a good fit. Stapper~et~al\@.~\cite{Stapper_1980} developed a yield prediction model for semiconductor memory chips by modeling the overdispersed distribution of number of faults per chip using the gamma-Poisson distribution which is closely related to the beta-binomial distribution.

For the beta-binomial channel model, we model the variables $\kcount{m}{0}$ and $\kcount{N-m}{1}$ as being distributed according to the beta-binomial distribution i.e.,
\begin{IEEEeqnarray}{rCl}
    p & \sim & \textrm{Beta}(a, b) \nonumber \\
    \kcount{m}{0}~|~p & \sim & \textrm{Binomial}(m, p) \nonumber \\
    \kcount{m}{0} & \sim & \textrm{Beta-Binomial}(m, a, b) \\
    q & \sim & \textrm{Beta}(c, d)  \nonumber \\
    \kcount{N-m}{1}~|~q & \sim & \textrm{Binomial}(N-m, q)  \nonumber \\
    \kcount{N-m}{1} & \sim & \textrm{Beta-Binomial}(N-m, c, d) \\
    \kcount{m}{0} & \indep & \kcount{N-m}{1} \label{eqn:bbm_xy_indep}
\end{IEEEeqnarray}
where $(a, b)$ and $(c, d)$ correspond to the parameters of a beta probability distribution defined as
\begin{IEEEeqnarray}{rCl}
    f(\theta; \alpha, \beta) & = & \frac{\theta^{\alpha-1} (1 - \theta)^{\beta-1}}{B(\alpha, \beta)}~~0 \leq \theta \leq 1\\
    B(\alpha, \beta) & = & \int_{0}^{1} \theta^{\alpha-1} (1 - \theta)^{\beta-1} \textrm{d}\theta
\end{IEEEeqnarray}
where $B(\alpha, \beta)$ represents the beta function.
Thus the Beta-Binomial (BBM) channel model is derived from a BAC model where the bit error probabilities $p$ and $q$ are random variables which vary from frame to frame and are distributed according to the beta distribution. The BBM channel model is a 4-parameter model (compared to the 2-parameter BAC) and hence the 2-BBM channel model for MLC flash memories will be an 8-parameter model.
The beta-binomial probability distributions of $\kcount{m}{0}$ and $\kcount{N-m}{1}$ are given by
\ifCLASSOPTIONonecolumn
\begin{IEEEeqnarray}{rCl}
    \Pr(\kcount{m}{0} = k) & = & {m \choose k} \frac{B(a+k, b+m-k)}{B(a, b)} \label{eqn:bbm_km0_dist}  \\
	\Pr(\kcount{N-m}{1} = k) & = & {N-m \choose k} \frac{B(c+k, d+N-m-k)}{B(c, d)}. \label{eqn:bbm_km1_dist}
\end{IEEEeqnarray}
\else
\begin{IEEEeqnarray}{rCl}
	\Pr(\kcount{m}{0} = k) & = & {m \choose k} \frac{B(a+k, b+m-k)}{B(a, b)} \label{eqn:bbm_km0_dist}
\end{IEEEeqnarray}
\begin{IEEEeqnarray}{rCl}
	\Pr(\kcount{N-m}{1} = k) & = & {N-m \choose k} \nonumber \\
	& & \frac{B(c+k, d+N-m-k)}{B(c, d)}. \label{eqn:bbm_km1_dist}
\end{IEEEeqnarray}
\fi
The mean and the variance of $\kcount{m}{0}$ and $\kcount{N-m}{1}$ are given by
    \begin{IEEEeqnarray}{rCl}
        \E[\kcount{m}{0}] & = & \frac{ma}{a+b}  \label{eqn:xm_bbm_mean} \\
    	\Var[\kcount{m}{0}] & = & \frac{mab(a+b+m)}{(a+b)^2(a+b+1)} \label{eqn:xm_bbm_var} \\
        \E[\kcount{N-m}{1}] & = & \frac{(N-m)c}{c+d} \label{eqn:ynm_bbm_mean} \\
    	\Var[\kcount{N-m}{1}] & = & \frac{(N-m)cd(c+d+N-m)}{(c+d)^2(c+d+1)}. \label{eqn:ynm_bbm_var}
    \end{IEEEeqnarray}
%
\ifCLASSOPTIONonecolumn
\begin{IEEEprop}
	The mean and the variance of $K$ for a BBM channel model are given by
	\begin{IEEEeqnarray}{rCl}
	    \E[K] & = & \frac{N}{2}\Bigg(\frac{a}{a+b} + \frac{c}{c+d} \Bigg) \label{eqn:mean_bbm} \\
        \Var[K] &=& \frac{N}{4}\Bigg(\frac{a(a+b)(a+2b+1) + Nab}{(a+b)^2(a+b+1)}\Bigg) + \frac{N}{4}\Bigg(\frac{c(c+d)(c+2d+1) + Ncd}{(c+d)^2(c+d+1)}\Bigg) \nonumber \\
                & & - \frac{N}{4}\Bigg(\frac{2ac}{(a+b)(c+d)} \Bigg). \label{eqn:var_bbm}
	\end{IEEEeqnarray}
    \label{prop:bbm_k_mean_var}
\end{IEEEprop}
\else
\begin{IEEEprop}
	The mean and the variance of $K$ for a BBM channel model are given by
	\begin{IEEEeqnarray}{rCl}
	    \E[K] & = & \frac{N}{2}\Bigg(\frac{a}{a+b} + \frac{c}{c+d} \Bigg) \label{eqn:mean_bbm} \\
        \Var[K] &=& \frac{N}{4}\Bigg(\frac{a(a+b)(a+2b+1) + Nab}{(a+b)^2(a+b+1)}\Bigg) + \nonumber \\
                & & \frac{N}{4}\Bigg(\frac{c(c+d)(c+2d+1) + Ncd}{(c+d)^2(c+d+1)}\Bigg) - \nonumber \\
                & & \frac{N}{4}\Bigg(\frac{2ac}{(a+b)(c+d)} \Bigg). \label{eqn:var_bbm}
	\end{IEEEeqnarray}
    \label{prop:bbm_k_mean_var}
\end{IEEEprop}
\fi
\begin{IEEEproof}
	See Appendix~\ref{app:proof_bbm_k_mean_var}.
\end{IEEEproof}
\begin{IEEEprop}
	The mean and the second moment of $K^{(0)}$ and $K^{(1)}$ for a BBM channel model are given by
	\begin{IEEEeqnarray}{rCl}
		\E[K^{(0)}] & = & \frac{N}{2}\Bigg(\frac{a}{a+b}\Bigg) \label{eqn:bbm_k0_mean}\\
		\E[(K^{(0)})^{2}] & = & \frac{N}{4}\Bigg(\frac{a(a+2b+1) + Na(a+1)}{(a+b)(a+b+1)}\Bigg) \label{eqn:bbm_k0_second_moment}\\
		\E[K^{(1)}] & = & \frac{N}{2}\Bigg(\frac{c}{c+d}\Bigg)  \label{eqn:bbm_k1_mean}\\
		\E[(K^{(1)})^{2}] & = & \frac{N}{4}\Bigg(\frac{c(c+2d+1) + Nc(c+1)}{(c+d)(c+d+1)}\Bigg). \label{eqn:bbm_k1_second_moment}
	\end{IEEEeqnarray}
    \label{prop:bbm_k0_k1_mean_second_moment}
\end{IEEEprop}
\begin{IEEEproof}
	See Appendix~\ref{app:proof_bbm_k0_k1_mean_second_moment}.
\end{IEEEproof}

The parameters $a$, $b$, $c$, $d$ of the BBM channel model are estimated from the sample moments of $K^{(0)}$ and $K^{(1)}$ using the method of moments~\cite{Skellam_1948}.
From P/E cycling experiment data, we obtain the sample mean and sample second moment estimates of the random variables $K^{(0)}$ and $K^{(1)}$ which represent the total number of $\ZOerror$ and $\OZerror$ bit errors per frame. Let $\mu_{1}$, $\mu_{2}$ represent the first and second moment estimates of $K^{(0)}$ and $\mu_{3}$, $\mu_{4}$ represent the first and second moment estimates of $K^{(1)}$. Solving the equations in Proposition~\ref{prop:bbm_k0_k1_mean_second_moment} for $a$, $b$, $c$, $d$, we have the parameter estimates
\ifCLASSOPTIONonecolumn
    \begin{IEEEeqnarray}{rClrCl}
        \hat{a} & = & \frac{\mu_{1}^{2}(N + 1) - 2\mu_{1}\mu_{2}}{N(\mu_2 - \mu_1) - \mu_{1}^{2}(N-1)}
        ~~~~~~ & \hat{b} & = & \hat{a}\Bigg(\frac{N}{2\mu_1} - 1 \Bigg) \\
        \hat{c} & = & \frac{\mu_{3}^{2}(N + 1) - 2\mu_{3}\mu_{4}}{N(\mu_4 - \mu_3) - \mu_{3}^{2}(N-1)}
        ~~~~~~ & \hat{d} & = & \hat{c}\Bigg(\frac{N}{2\mu_3} - 1 \Bigg).
    \end{IEEEeqnarray}
\else
    \begin{IEEEeqnarray}{rClrCl}
        \hat{a} & = & \frac{\mu_{1}^{2}(N + 1) - 2\mu_{1}\mu_{2}}{N(\mu_2 - \mu_1) - \mu_{1}^{2}(N-1)}
        ~~~~ & \hat{b} & = & \hat{a}\Bigg(\frac{N}{2\mu_1} - 1 \Bigg) \\
        \hat{c} & = & \frac{\mu_{3}^{2}(N + 1) - 2\mu_{3}\mu_{4}}{N(\mu_4 - \mu_3) - \mu_{3}^{2}(N-1)}
        ~~~~ & \hat{d} & = & \hat{c}\Bigg(\frac{N}{2\mu_3} - 1 \Bigg).
    \end{IEEEeqnarray}
\fi
An algorithmic description of the BBM channel model is presented in~Algorithm~\ref{algo:bbm_channel_model}.
\begin{algorithm}[!ht]
	\caption{\mbox{BBM} Channel Model Implementation}
	\label{algo:bbm_channel_model}
	\algsetup{
		linenosize=\small
	}
	\begin{algorithmic}[1]
        \REQUIRE Input frame $\mathbf{x}$ of length $N$, BBM channel model parameters $(a$, $b$, $c$, $d)$.
        \ENSURE Data frame with errors $\mathbf{y}$.
        \ifCLASSOPTIONonecolumn
            \STATE Generate two independent random samples, $p~\sim~\textrm{Beta}(a, b)$ and $q~\sim~\textrm{Beta}(c, d)$.
        \else
            \STATE Generate two independent random samples, \\ $p~\sim~\textrm{Beta}(a, b)$ and $q~\sim~\textrm{Beta}(c, d)$.
        \fi
        \STATE $\mathbf{y} = \textrm{BAC}(\mathbf{x}, p, q)$ [Use Algorithm~\ref{algo:bac_channel_model}].
    \end{algorithmic}
\end{algorithm}
\begin{table}[!ht]
    \caption{Upper page BBM channel model parameter estimates for vendor-A and vendor-B chips. $N = 8192$.}
	\label{table:bbm_parameter_table}
	\centering
	\bgroup
	\ifCLASSOPTIONonecolumn
    	\begin{tabular}{|c|c c c c|c c c c|}
    		\hline
    		\textbf{P/E Cycles} & \multicolumn{4}{|c|}{\textbf{Vendor-A}} &  \multicolumn{4}{|c|}{\textbf{Vendor-B}} \\
            \hline
    		& a & b & c & d & a & b & c & d \\
    		\hline
            2000 & 12.72 & 46368.34 & 8.05 & 42569.08 & 10.82 & 302596.64 & 6.86 & 43747.02 \\
            \hline
            4000 & 25.95 & 20940.98 & 15.46 & 23556.92 & 11.39 & 48028.59 & 6.00 & 13142.88 \\
            \hline
    		6000 & 22.67 & 7596.71 & 18.16 & 11890.14 & 15.58 & 20535.47 & 7.16 & 7193.92 \\
            \hline
    		8000 & 20.72 & 4143.52 & 22.28 & 7821.13 & 15.28 & 9068.43 & 7.58 & 4092.87 \\
            \hline
    		10000 & 21.36 & 2819.03 & 26.12 & 5890.35 & 13.36 & 4142.23 & 9.28 & 2938.88 \\
    		\hline
    	\end{tabular}
	\else
		\def\arraystretch{1.5}
        \tabcolsep=0.11cm
        \resizebox{0.98\columnwidth}{!}{
        \begin{tabular}{|c|c c c c|c c c c|}
            \hline
            \textbf{P/E Cycles} & \multicolumn{4}{|c|}{\textbf{Vendor-A}} &  \multicolumn{4}{|c|}{\textbf{Vendor-B}} \\
            \hline
            & a & b & c & d & a & b & c & d \\
            \hline
            2000 & 12.72 & 46368.34 & 8.05 & 42569.08 & 10.82 & 302596.64 & 6.86 & 43747.02 \\
            \hline
            4000 & 25.95 & 20940.98 & 15.46 & 23556.92 & 11.39 & 48028.59 & 6.00 & 13142.88 \\
            \hline
            6000 & 22.67 & 7596.71 & 18.16 & 11890.14 & 15.58 & 20535.47 & 7.16 & 7193.92 \\
            \hline
            8000 & 20.72 & 4143.52 & 22.28 & 7821.13 & 15.28 & 9068.43 & 7.58 & 4092.87 \\
            \hline
            10000 & 21.36 & 2819.03 & 26.12 & 5890.35 & 13.36 & 4142.23 & 9.28 & 2938.88 \\
            \hline
        \end{tabular}
        }
	\fi
	\egroup
\end{table}

For evaluation of the BBM channel model, we compute $\E[K]$ and $\Var[K]$ using Proposition~\ref{prop:bbm_k_mean_var}. 
Corresponding to the example used for evaluating the BAC model, the parameter estimates of the upper page BBM channel model for vendor-A are as shown in Table~\ref{table:bbm_parameter_table} and using these parameter estimates, we obtain $\E[K] = 32.01$ and $\Var[K] = 57.88$ for $N = 8192$ at 8,000 P/E cycles. Comparing with the results from Table~\ref{table:sample_mean_var_table},
we observe that the $\Var[K]$ obtained using the BBM channel model is still lower than the sample variance; however, it is clear that the BBM channel model is vastly better at modeling the overdispersed number of bit errors per frame empirical data than the BAC model. This will be even more evident based on the ECC FER performance estimation results presented in~Section~\ref{sec:empirical_results}.
\ifCLASSOPTIONonecolumn
\begin{figure}
    \begin{minipage}[b]{0.49\linewidth}
        \centering
        \includegraphics[width=\textwidth]{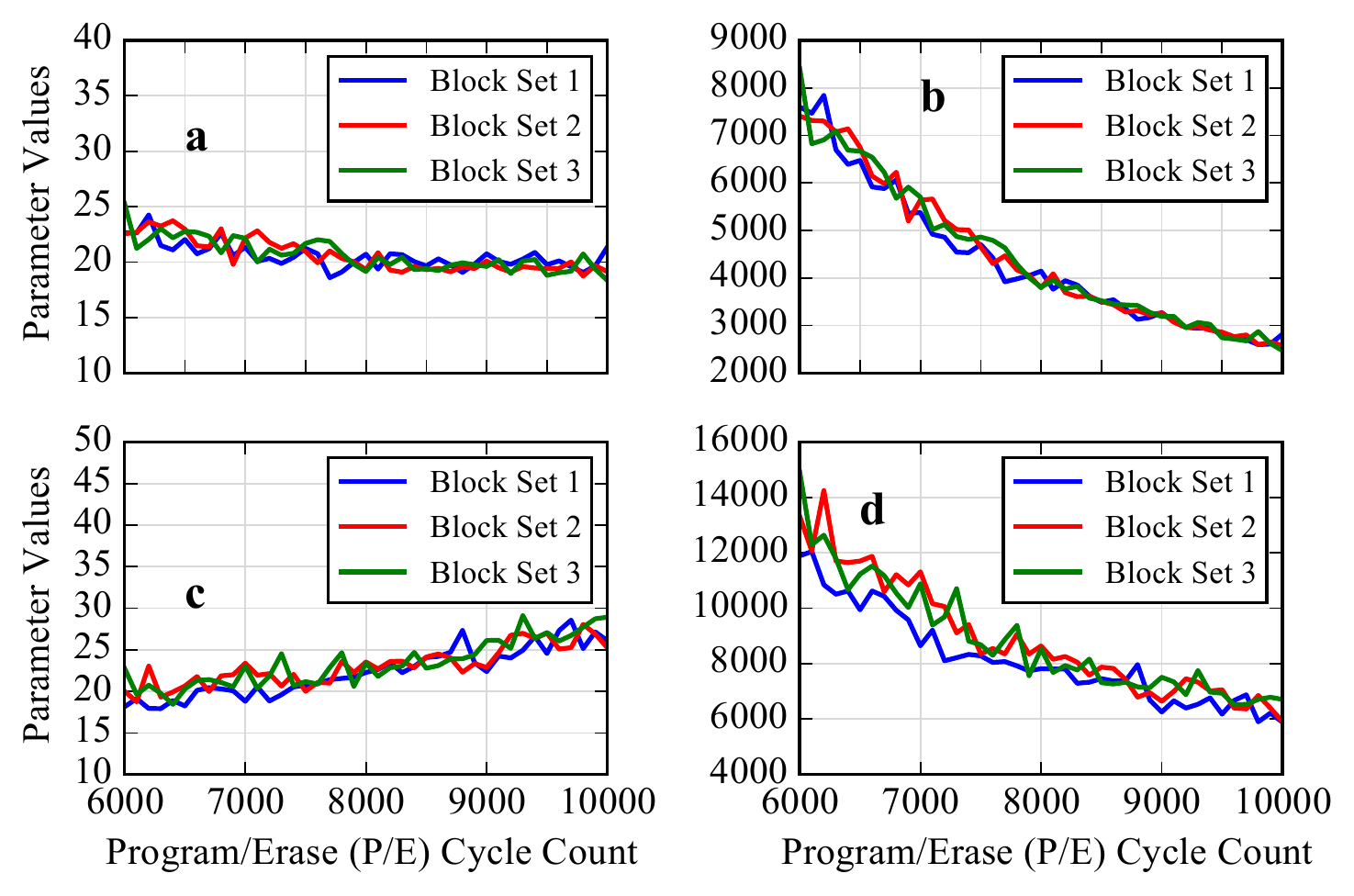}
    	\caption{Variation of parameter estimates for the upper page BBM channel model (($a$, $b$) for $\ZOerror$ error, ($c$, $d$) for $\OZerror$ error) for 3 different 4-block sets for \mbox{vendor-A} chip. $N = 8192$.}
    	\label{fig:up_parameters_vendor_a}
    \end{minipage}
    \hspace{0.2cm}
    \begin{minipage}[b]{0.49\linewidth}
        \centering
    	\includegraphics[width=\textwidth]{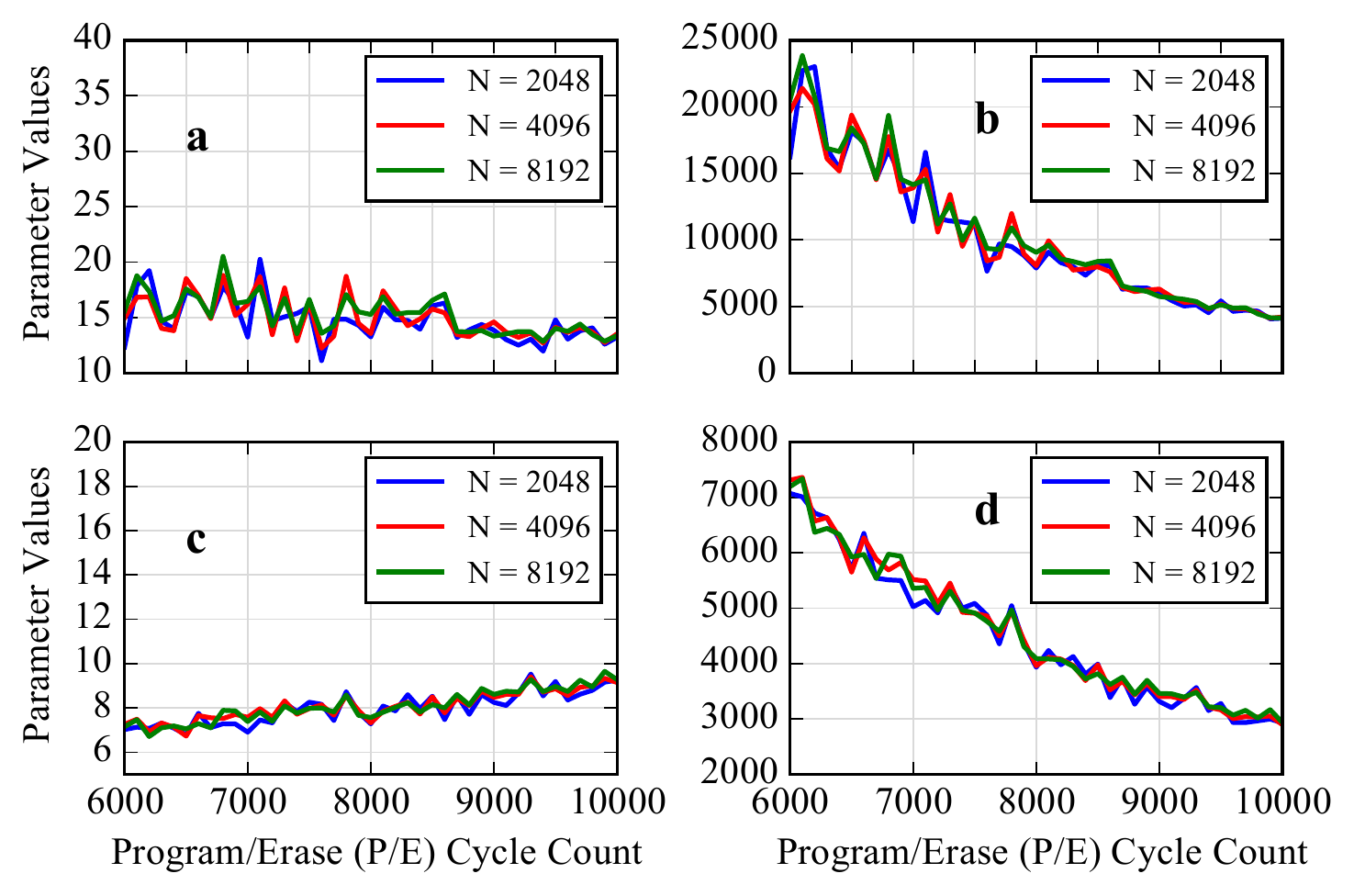}
    	\caption{Variation of parameter estimates for the upper page BBM channel model (($a$, $b$) for $\ZOerror$ error, ($c$, $d$) for $\OZerror$ error) for 3 different frame lengths for \mbox{vendor-B} chip.}
    	\label{fig:up_parameters_frame_length_vendor_b}
    \end{minipage}
\end{figure}
\else
    \begin{figure}
    	\centering
        \includegraphics[width=0.49\textwidth]{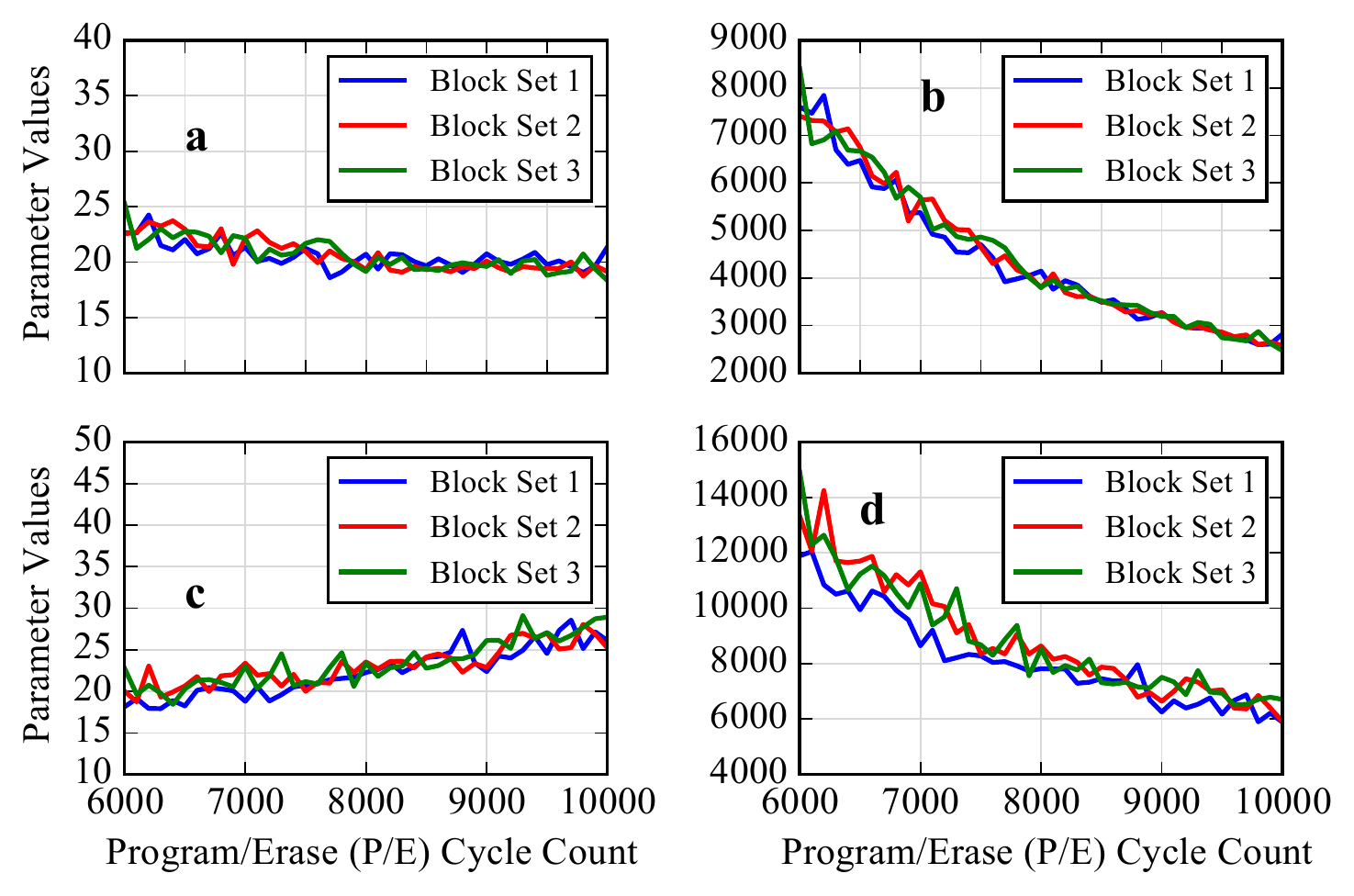}
    	\caption{Variation of parameter estimates for the upper page BBM channel model (($a$, $b$) for $\ZOerror$ error, ($c$, $d$) for $\OZerror$ error) for 3 different 4-block sets for \mbox{vendor-A} chip. $N = 8192$.}
    	\label{fig:up_parameters_vendor_a}
    \end{figure}
    \begin{figure}
    	\centering
        \includegraphics[width=0.49\textwidth]{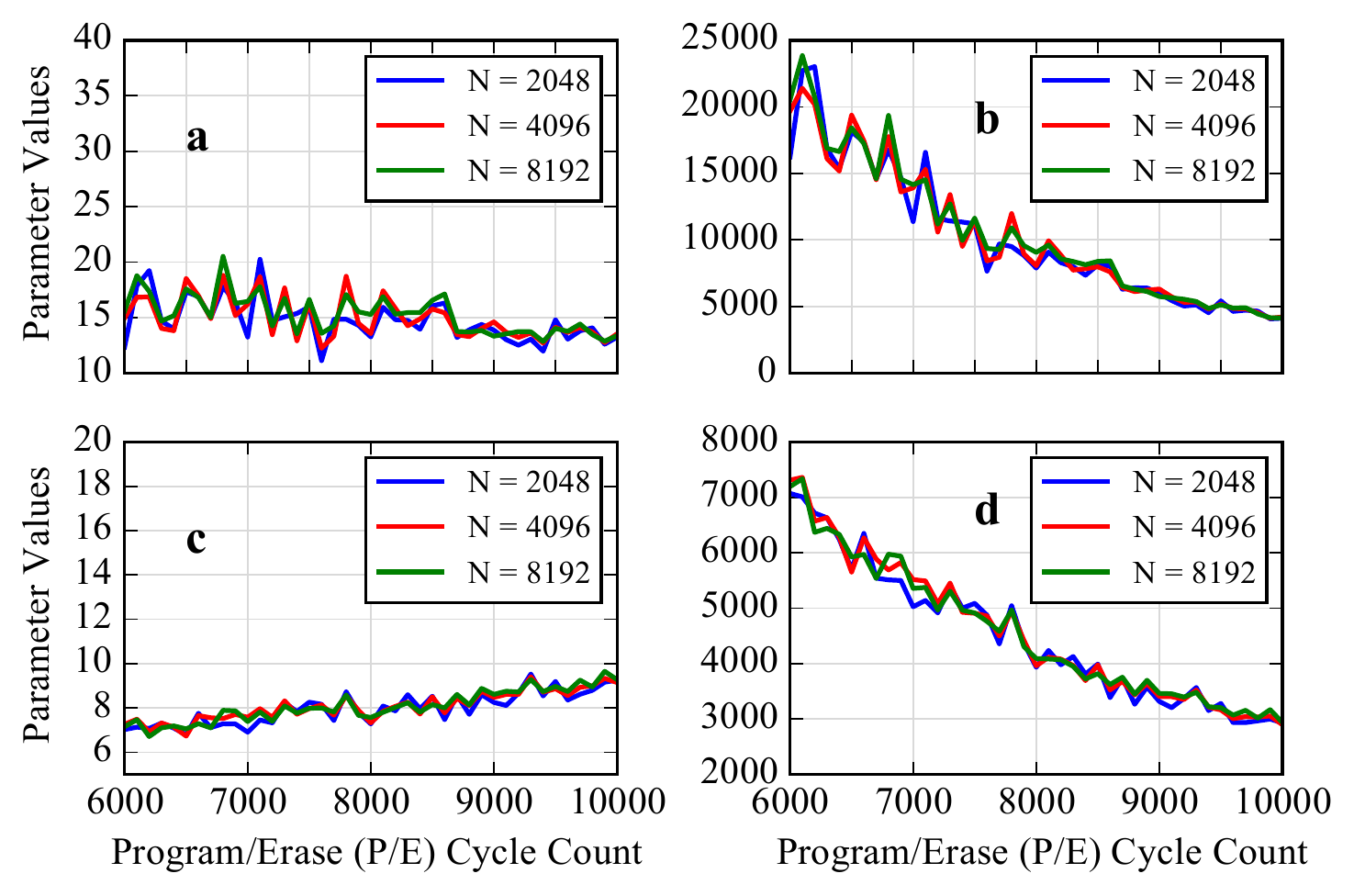}
    	\caption{Variation of parameter estimates for the upper page BBM channel model (($a$, $b$) for $\ZOerror$ error, ($c$, $d$) for $\OZerror$ error) for 3 different frame lengths for \mbox{vendor-B} chip.}
    	\label{fig:up_parameters_frame_length_vendor_b}
    \end{figure}
\fi

We also observe remarkable consistency in the parameter estimates of the BBM channel model across different blocks of the same MLC flash memory chip. Fig.~\ref{fig:up_parameters_vendor_a} shows the empirical parameter estimates corresponding to the upper page BBM channel models for vendor-A chip using data collected from 3 different sets of 4 contiguous blocks of the MLC flash memory chip. Fig.~\ref{fig:up_parameters_frame_length_vendor_b} shows the empirical parameter estimates corresponding to the upper page BBM channel models for vendor-B chip obtained using different frame sizes. Although not shown (due to lack of space), we also observe similar consistency in the lower page parameter estimates for both the vendor chips using different sets of blocks on the same chip and different frame sizes. We also note that the estimates for lower page parameters $a$ and $b$ will be noisy because the $\ZOerror$ bit error rate in the lower page is extremely small.
This consistency suggests that we may be able to model every flash memory chip with just 8 parameters of the 2-BBM channel model for accurate ECC FER performance estimation.

\subsection{Normal and Poisson Approximation Channel Models}
To model the overdispersed number of bit errors per frame empirical data, an alternative approach from a statistical viewpoint is to consider approximations to the binomial probability distribution which retain the general shape of the binomial distribution and whose mean and variance can be controlled independently. We propose two such channel models for MLC flash memories based on the normal and Poisson probability distributions called the 2-Normal Approximation to the BAC (\mbox{2-NA-BAC}) model and the 2-Poisson Approximation to the BAC (\mbox{2-PA-BAC}) model respectively. Similar to the \mbox{2-BAC} and \mbox{2-BBM} channel models, the \mbox{2-NA-BAC} (resp., \mbox{2-PA-BAC}) model consists of two independent \mbox{NA-BAC} (resp., \mbox{PA-BAC}) models for the lower and upper pages of MLC flash memories. The design goal for the NA-BAC and PA-BAC models is to ensure a match between the mean and variance statistics of the data from the model and the observed sample mean and sample variance. Based on this, we define rules for the normal and Poisson approximation as follows.

Let $\mu_0$ and $\sigma_0^{2}$ denote the sample mean and sample variance of $K^{(0)}$ and $\mu_1$ and $\sigma_1^{2}$ denote the sample mean and sample variance of $K^{(1)}$. Let $\mathcal{N}(\mu, \sigma^2)$ denote a normal distribution with mean $\mu$ and variance $\sigma^2$ and let $\mathcal{P}(\lambda)$ denote a Poisson distribution with rate parameter $\lambda$. Let $g_0$
and $g_1$ represent the sampled number of $\ZOerror$ and $\OZerror$ bit errors per frame.
\begin{IEEEdefinition}
The normal approximation rules for the \mbox{NA-BAC} model are given by
\ifCLASSOPTIONonecolumn
    \begin{IEEEeqnarray}{rClrCl}
        g_0 & = & [\hat{g}_0]~\textrm{where}~\hat{g}_0~\sim~\mathcal{N}\big(\mu_0, \sigma_0^{2}\big) ~~~~~~~~~~~~ & g_1 & = & [\hat{g}_1]~\textrm{where}~\hat{g}_1~\sim~\mathcal{N}\big(\mu_1, \sigma_1^{2}\big).
    \end{IEEEeqnarray}
\else
    \begin{IEEEeqnarray}{rCl}
        g_0 & = & [\hat{g}_0]~\textrm{where}~\hat{g}_0~\sim~\mathcal{N}\big(\mu_0, \sigma_0^{2}\big) \nonumber \\
        g_1 & = & [\hat{g}_1]~\textrm{where}~\hat{g}_1~\sim~\mathcal{N}\big(\mu_1, \sigma_1^{2}\big).
    \end{IEEEeqnarray}
\fi
where $[\cdot]$ denotes the round to nearest integer operator.
\label{def:nabac_rules}
\end{IEEEdefinition}
\begin{IEEEdefinition}
The Poisson approximation rules for the \mbox{PA-BAC} model are given by
\ifCLASSOPTIONonecolumn
    \begin{IEEEeqnarray}{rClrCl}
         g_0 & = & \hat{g}_0 - (\sigma_0^{2} - \mu_0)~\textrm{where}~\hat{g}_0~\sim~ \mathcal{P}\big(\sigma_0^{2}\big)  ~~~~ &
         g_1 & = & \hat{g}_1 - (\sigma_1^{2} - \mu_1)~\textrm{where}~\hat{g}_1~\sim~ \mathcal{P}\big(\sigma_1^{2}\big).
    \end{IEEEeqnarray}
\else
    \begin{IEEEeqnarray}{rCl}
         g_0 & = & \hat{g}_0 - (\sigma_0^{2} - \mu_0)~\textrm{where}~\hat{g}_0~\sim~ \mathcal{P}\big(\sigma_0^{2}\big)
         \nonumber \\
         g_1 & = & \hat{g}_1 - (\sigma_1^{2} - \mu_1)~\textrm{where}~\hat{g}_1~\sim~ \mathcal{P}\big(\sigma_1^{2}\big).
    \end{IEEEeqnarray}
\fi
\label{def:pabac_rules}
\end{IEEEdefinition}
\begin{algorithm}
	\caption{\mbox{NA-BAC} and \mbox{PA-BAC} Model Implementation}
	\label{algo:normal_poisson_approx_channel_model}
	\algsetup{
		linenosize=\small
	}
	\begin{algorithmic}[1]
        \REQUIRE Input frame $\mathbf{x}$ of length $N$, sample $(\E[K^{(0)}], \Var[K^{(0)}])$, and sample $(\E[K^{(1)}], \Var[K^{(1)}])$. 
        \ENSURE Data frame with errors $\mathbf{y}$.
        \STATE Generate integers $g_0$, $g_1$ according to the Normal or Poisson approximation rules.
        \STATE $\mathcal{T}_{0} = \{ i~|~x_i = 0 \}$, $\mathcal{T}_{1} = \{ i~|~x_i = 1 \}$.
        \STATE Pick subsets $\mathcal{E}_{0}$ of size $g_0$ and $\mathcal{E}_{1}$ of size $g_1$ uniformly at random from $\mathcal{T}_{0}$ and $\mathcal{T}_{1}$, respectively.
        \STATE Create a binary error vector $\mathbf{e}$ of length $N$ such that $e_{i} = 1$ if $i \in \mathcal{E}_{0} \cup \mathcal{E}_{1}$.
        \STATE $\mathbf{y} = \mathbf{x} \oplus \mathbf{e}$.
    \end{algorithmic}
\end{algorithm}
Based on these rules, an algorithmic description of the \mbox{NA-BAC} and \mbox{PA-BAC} models is presented in~Algorithm~\ref{algo:normal_poisson_approx_channel_model}.
The normal probability distribution is a continuous distribution with infinite support whereas the variables $K^{(0)}$ and $K^{(1)}$ being modeled have finite support and are discrete (integers). Hence we require the round to nearest integer function in Definition~\ref{def:nabac_rules}. The Poisson probability distribution is a discrete distribution with an infinite support set.
Using goodness of fit tests in~Section~\ref{sec:empirical_results}, we show that the \mbox{2-NA-BAC} and \mbox{2-PA-BAC} models are a better fit than the 2-BAC model for the observed empirical data. However,
the \mbox{2-NA-BAC} and the \mbox{2-PA-BAC} models are not as good a fit as the \mbox{2-BBM} model to describe the bit errors in MLC flash memories.

%% file: sections/ecc_performance_prediction.tex
\section{Simulation Results and Evaluation of Channel Models}
\label{sec:empirical_results}

In this section, we provide a quantitative evaluation of the proposed channel models for MLC flash memories. For this we consider two viewpoints. The first one is a purely statistical viewpoint where we perform the Kolmogorov-Smirnov \mbox{(K-S)} Two Sample test~\cite{Massey_1951} to evaluate the goodness of fit of the proposed channel models when compared with the empirical data. Next, we evaluate the proposed channel models for their application in ECC FER performance estimation. We emphasize the results of this latter evaluation when compared to the former, as accurate ECC FER performance estimation has been the main driving factor in the design of the proposed channel models.

\subsection{Statistical Goodness of Fit Tests}
The Kolmogorov-Smirnov (\mbox{K-S}) Two Sample test is a commonly used statistical test for determining if two sets of data samples are drawn from the same probability distribution. The \mbox{K-S} test is a very general test in that it makes no assumptions about the underlying probability distributions of the input data samples and is a non-parametric test~\cite{Massey_1951}. This makes it suitable for our purpose as we have a varied set of underlying probability distributions of the number of bit errors per frame corresponding to the proposed channel models. The BAC and BBM model distributions do not match any well known probability distributions exactly although, they are close to the binomial distribution, and the \mbox{NA-BAC} and \mbox{PA-BAC} model distributions are approximately normal and Poisson respectively.

We perform K-S Two Sample tests comparing the number of bit errors per frame data samples from the proposed channel models to the empirical data obtained from P/E cycling experiments. The empirical data sample sizes, i.e., number of frames for each page, are $8704$ for \mbox{vendor-A} and $4096$ for \mbox{vendor-B}, respectively. For the BAC, BBM, \mbox{NA-BAC} and \mbox{PA-BAC} models, we simulate $10000$ frames. The beta random variates to simulate the BBM channel model and the K-S Two Sample test
statistic values are computed using the SciPy library~\cite{SciPy_2016}. The test statistic values are shown in~Tables~\ref{table:ks_test_stat_table} and \ref{table:ks_test_stat_table_4000} for $8,000$ and $4,000$ P/E cycles, respectively. The null hypothesis is that the data samples from a proposed channel model and empirical data belong to the same underlying probability distribution. The test statistic is indicative of the difference in underlying probability distributions of the two input data samples. From Table~\ref{table:ks_test_stat_table}, we see that the test statistic values are consistently low for the BBM channel model, thus indicating that it provides the best fit to the empirical data among all the proposed channel models. The p-values recorded (not shown) for all the K-S Two Sample tests in~Tables~\ref{table:ks_test_stat_table} and \ref{table:ks_test_stat_table_4000} are smaller than $0.01$ indicating that the test statistic values are estimated with a significant level of confidence. The K-S Two Sample test compares the cumulative distribution functions (CDF) obtained from input data samples to compute the test statistic. Fig.~\ref{fig:cdfs_8000_vendor_a} and Fig.~\ref{fig:cdfs_8000_vendor_b} provide a visual comparison of these CDFs corresponding to \mbox{vendor-A} and \mbox{vendor-B} chips.
\begin{table}[h]
	\caption{Test statistic values from K-S two sample tests comparing the lower and upper page BAC, BBM, \mbox{NA-BAC}, \mbox{PA-BAC} models with empirical data at 8,000 P/E cycles. Frame length $N = 8192$.}
	\label{table:ks_test_stat_table}
	\centering
	\bgroup
	\ifCLASSOPTIONonecolumn
		\begin{tabular}{|c|c c |c c|}
			\hline
			\textbf{K-S Two Sample Tests} & \multicolumn{2}{|c|}{\textbf{Vendor-A}} & \multicolumn{2}{|c|}{\textbf{Vendor-B}} \\
			\hhline{~----}
			 	& \textbf{Lower Page} & \textbf{Upper Page} & \textbf{Lower Page} & \textbf{Upper Page} \\
			\hline
			BAC vs. Experiment & 0.0979 & 0.0744 & 0.1278 & 0.0669 \\
			\hline
			BBM vs. Experiment & 0.0386 & 0.0357 & 0.0190 & 0.0135 \\
			\hline
			NA-BAC vs. Experiment & 0.0430 & 0.0715 & 0.0373 & 0.0659 \\
			\hline
			PA-BAC vs. Experiment & 0.0268 & 0.0777 & 0.0337 & 0.1008 \\
			\hline
		\end{tabular}
	\else
		\def\arraystretch{1.5}
		\tabcolsep=0.11cm
		\resizebox{0.98\columnwidth}{!}{
		\begin{tabular}{|c|c c |c c|}
			\hline
			\textbf{K-S Two Sample Tests} & \multicolumn{2}{|c|}{\textbf{Vendor-A}} & \multicolumn{2}{|c|}{\textbf{Vendor-B}} \\
			\hhline{~----}
			 	& \textbf{Lower Page} & \textbf{Upper Page} & \textbf{Lower Page} & \textbf{Upper Page} \\
			\hline
			BAC v/s Experiment & 0.0979 & 0.0744 & 0.1278 & 0.0669 \\
			\hline
			BBM v/s Experiment & 0.0386 & 0.0357 & 0.0190 & 0.0135 \\
			\hline
			NA-BAC v/s Experiment & 0.0430 & 0.0715 & 0.0373 & 0.0659 \\
			\hline
			PA-BAC v/s Experiment & 0.0268 & 0.0777 & 0.0337 & 0.1008 \\
			\hline
		\end{tabular}
		}
	\fi
	\egroup
\end{table}
\begin{table}[h]
	\caption{Test statistic values from K-S two sample tests comparing the lower and upper page BAC, BBM, \mbox{NA-BAC}, \mbox{PA-BAC} models with empirical data at 4,000 P/E cycles. Frame length $N = 8192$.}
	\label{table:ks_test_stat_table_4000}
	\centering
	\bgroup
	\ifCLASSOPTIONonecolumn
		\begin{tabular}{|c|c c |c c|}
			\hline
			\textbf{K-S Two Sample Tests} & \multicolumn{2}{|c|}{\textbf{Vendor-A}} & \multicolumn{2}{|c|}{\textbf{Vendor-B}} \\
			\hhline{~----}
			 	& \textbf{Lower Page} & \textbf{Upper Page} & \textbf{Lower Page} & \textbf{Upper Page} \\
			\hline
			BAC vs. Experiment & 0.0498 & 0.0291 & 0.0436 & 0.0422 \\
			\hline
			BBM vs. Experiment & 0.0268 & 0.0153 & 0.0137 & 0.0053 \\
			\hline
			NA-BAC vs. Experiment & 0.0575 & 0.0973 & 0.0632 & 0.1191 \\
			\hline
			PA-BAC vs. Experiment & 0.0642 & 0.0703 & 0.0223 & 0.1953 \\
			\hline
		\end{tabular}
	\else
		\def\arraystretch{1.5}
		\tabcolsep=0.11cm
		\resizebox{0.98\columnwidth}{!}{
		\begin{tabular}{|c|c c |c c|}
			\hline
			\textbf{K-S Two Sample Tests} & \multicolumn{2}{|c|}{\textbf{Vendor-A}} & \multicolumn{2}{|c|}{\textbf{Vendor-B}} \\
			\hhline{~----}
			 	& \textbf{Lower Page} & \textbf{Upper Page} & \textbf{Lower Page} & \textbf{Upper Page} \\
			\hline
			BAC vs. Experiment & 0.0498 & 0.0291 & 0.0436 & 0.0422 \\
			\hline
			BBM vs. Experiment & 0.0268 & 0.0153 & 0.0137 & 0.0053 \\
			\hline
			NA-BAC vs. Experiment & 0.0575 & 0.0973 & 0.0632 & 0.1191 \\
			\hline
			PA-BAC vs. Experiment & 0.0642 & 0.0703 & 0.0223 & 0.1953 \\
			\hline
		\end{tabular}
		}
	\fi
	\egroup
\end{table}
\ifCLASSOPTIONonecolumn
	\begin{figure}
		\centering
		\includegraphics[width=0.7\textwidth]{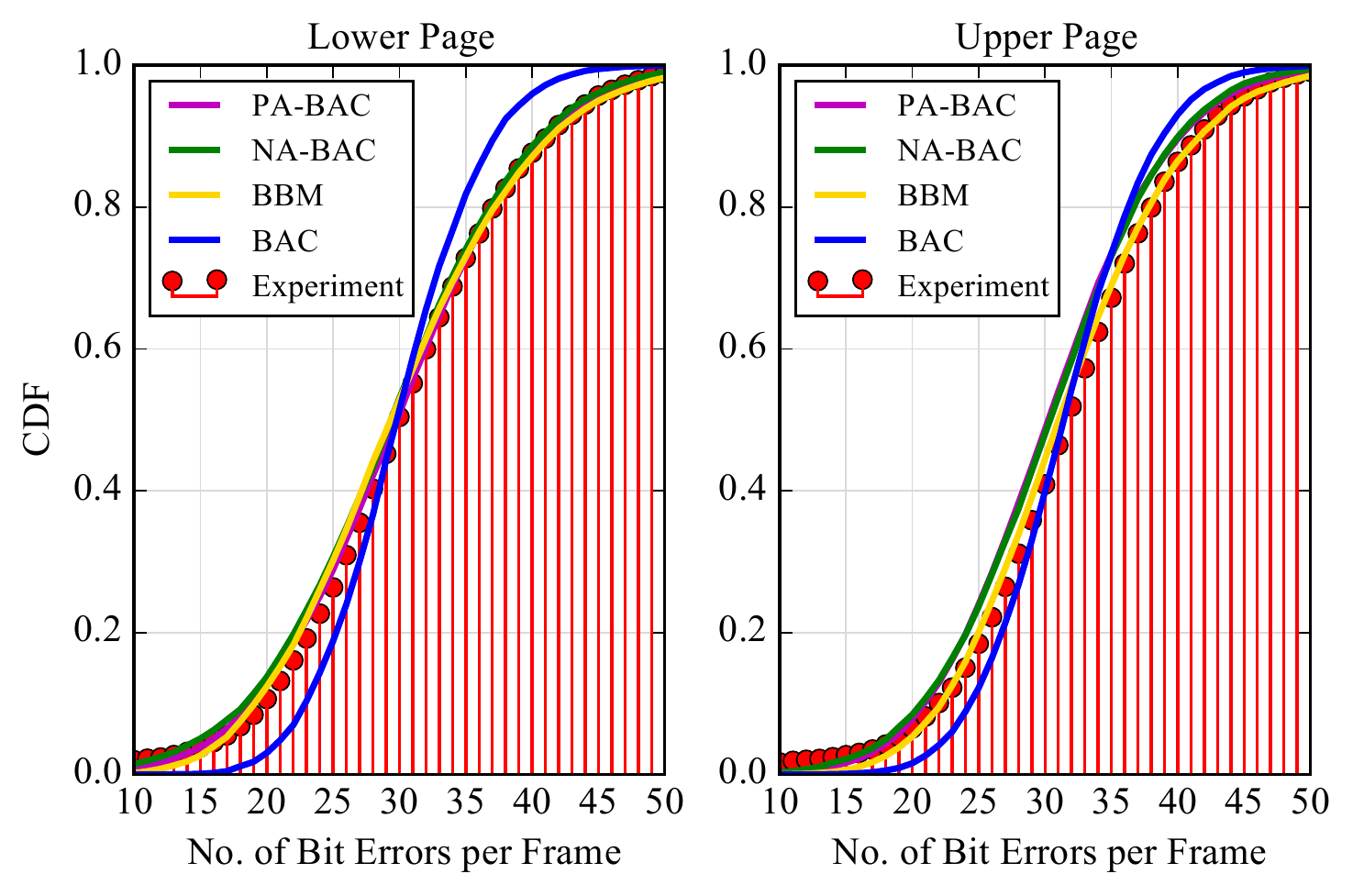}
		\caption{Comparison of CDFs for number of bit errors per frame observed from empirical data and from the BAC, BBM, NA-BAC, PA-BAC models at 8,000 P/E cycles for \mbox{vendor-A} chip.}
		\label{fig:cdfs_8000_vendor_a}
	\end{figure}
\else
	\begin{figure}
		\centering
		\includegraphics[width=0.47\textwidth]{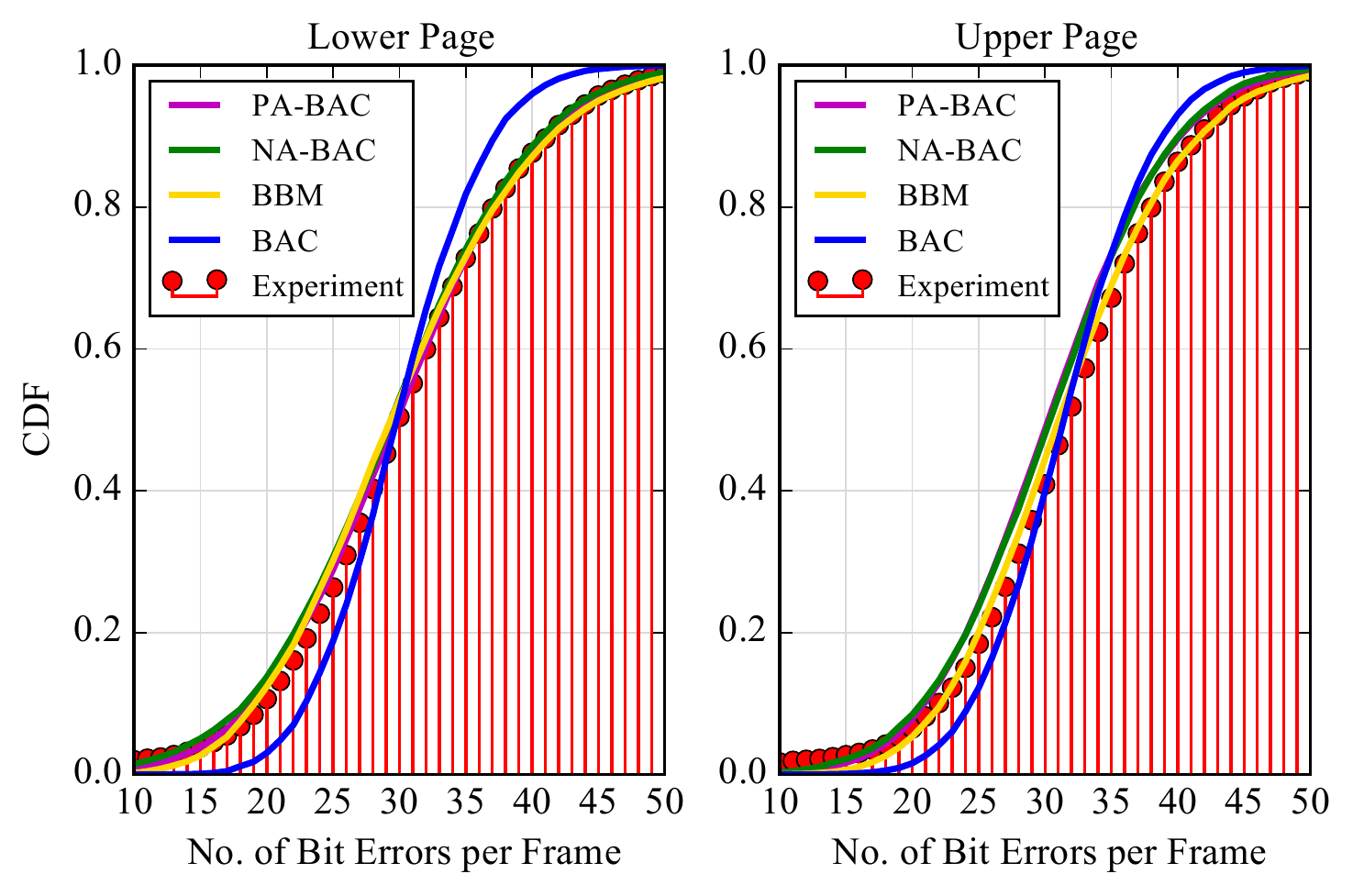}
		\caption{Comparison of CDFs for number of bit errors per frame observed from empirical data and from the BAC, BBM, NA-BAC, PA-BAC models at 8,000 P/E cycles for \mbox{vendor-A} chip.}
		\label{fig:cdfs_8000_vendor_a}
	\end{figure}
\fi
\ifCLASSOPTIONonecolumn
	\begin{figure}
		\centering
		\includegraphics[width=0.7\textwidth]{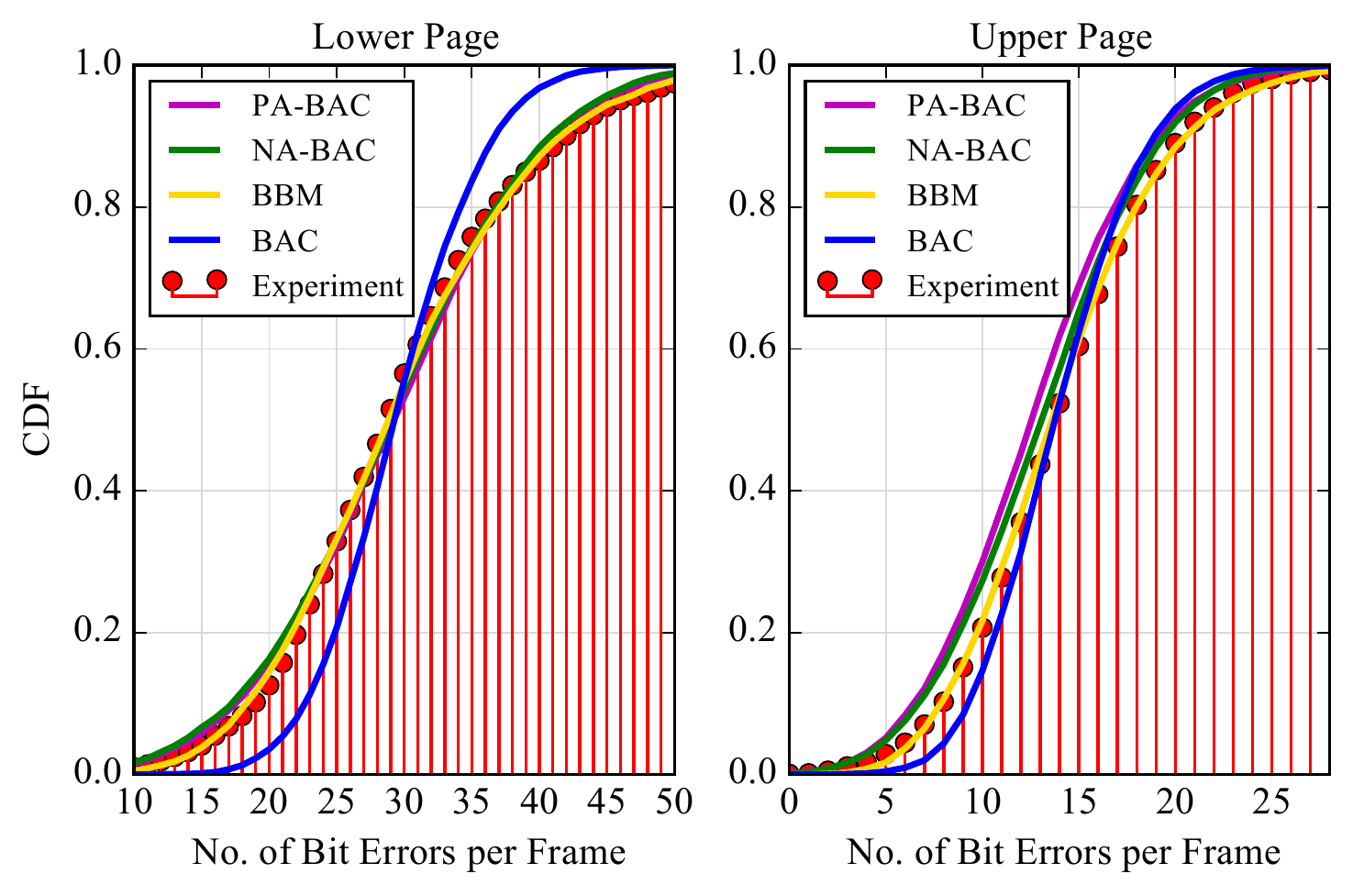}
		\caption{Comparison of CDFs for number of bit errors per frame observed from empirical data and from the BAC, BBM, NA-BAC, PA-BAC models at 8,000 P/E cycles for \mbox{vendor-B} chip.}
		\label{fig:cdfs_8000_vendor_b}
	\end{figure}
\else
	\begin{figure}
		\centering
		\includegraphics[width=0.47\textwidth]{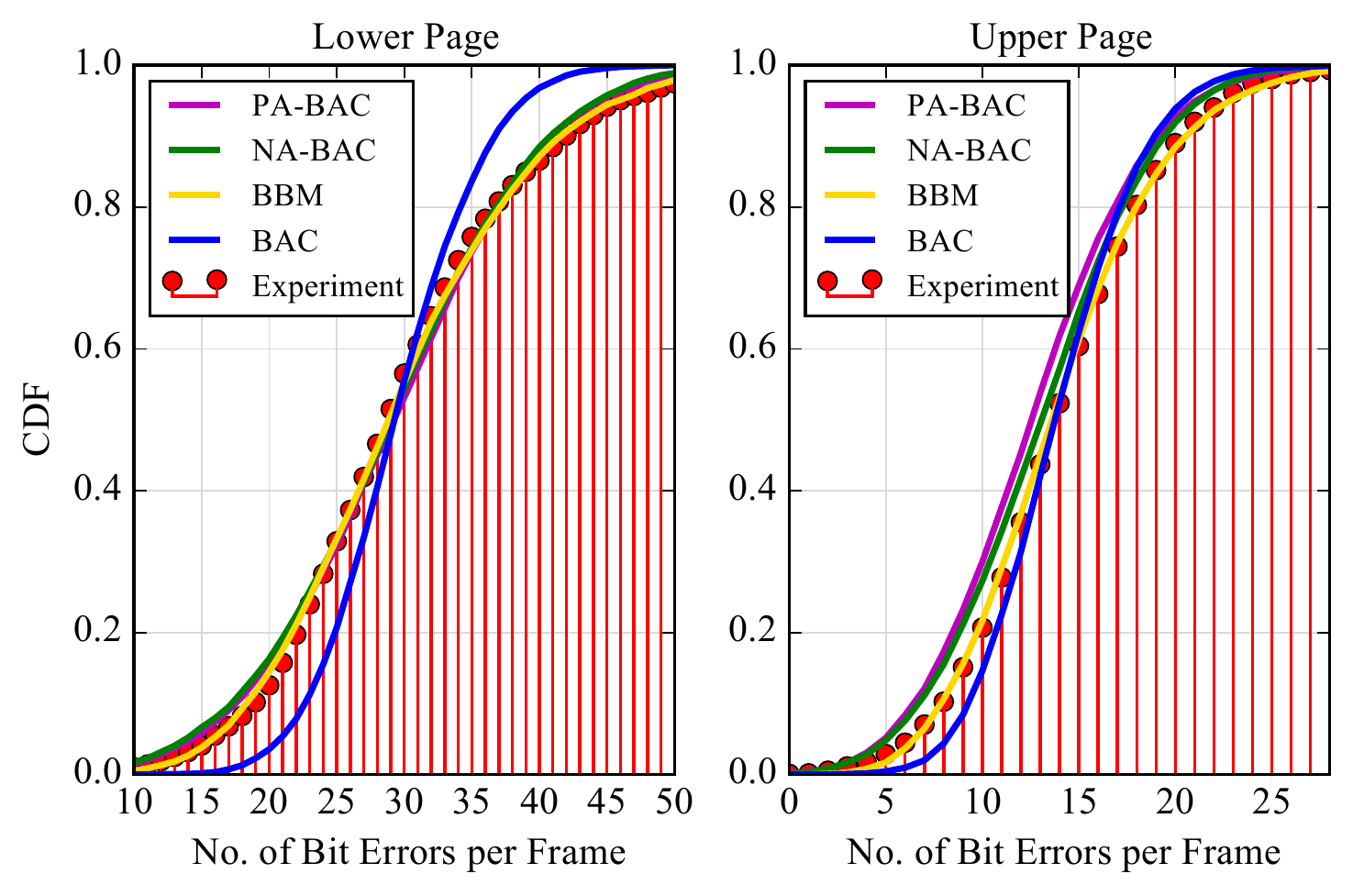}
		\caption{Comparison of CDFs for number of bit errors per frame observed from empirical data and from the BAC, BBM, NA-BAC, PA-BAC models at 8,000 P/E cycles for \mbox{vendor-B} chip.}
		\label{fig:cdfs_8000_vendor_b}
	\end{figure}
\fi
%
%
\subsection{ECC FER Performance Estimation}
We evaluate the proposed channel models for their accuracy in ECC FER performance estimation using binary BCH, LDPC, and polar codes. The choice of these ECCs reflects the fact that BCH and LDPC codes are already being used in practical flash memory applications, while polar codes are a promising candidate for the future. The baseline ECC FER performance estimates are obtained from the empirical error data collected from MLC flash memory chips during P/E cycling experiments. As pseudo-random data was written to the flash memory chips during P/E cycling experiments, for ECC decoding we assume an all-zero codeword as the transmitted codeword with the error vector obtained from the empirical error data. This assumption is valid because all the ECCs considered are linear codes. To estimate the ECC FER performance using the proposed channel models, Monte-Carlo simulations are used where pseudo-random codewords of the ECC are generated and transmitted through the appropriate channel model and the received codeword is decoded. At least 400 frame errors are recorded for FER estimation.
\ifCLASSOPTIONonecolumn
	\begin{figure}[h]
		\centering
		\includegraphics[width=0.7\textwidth]{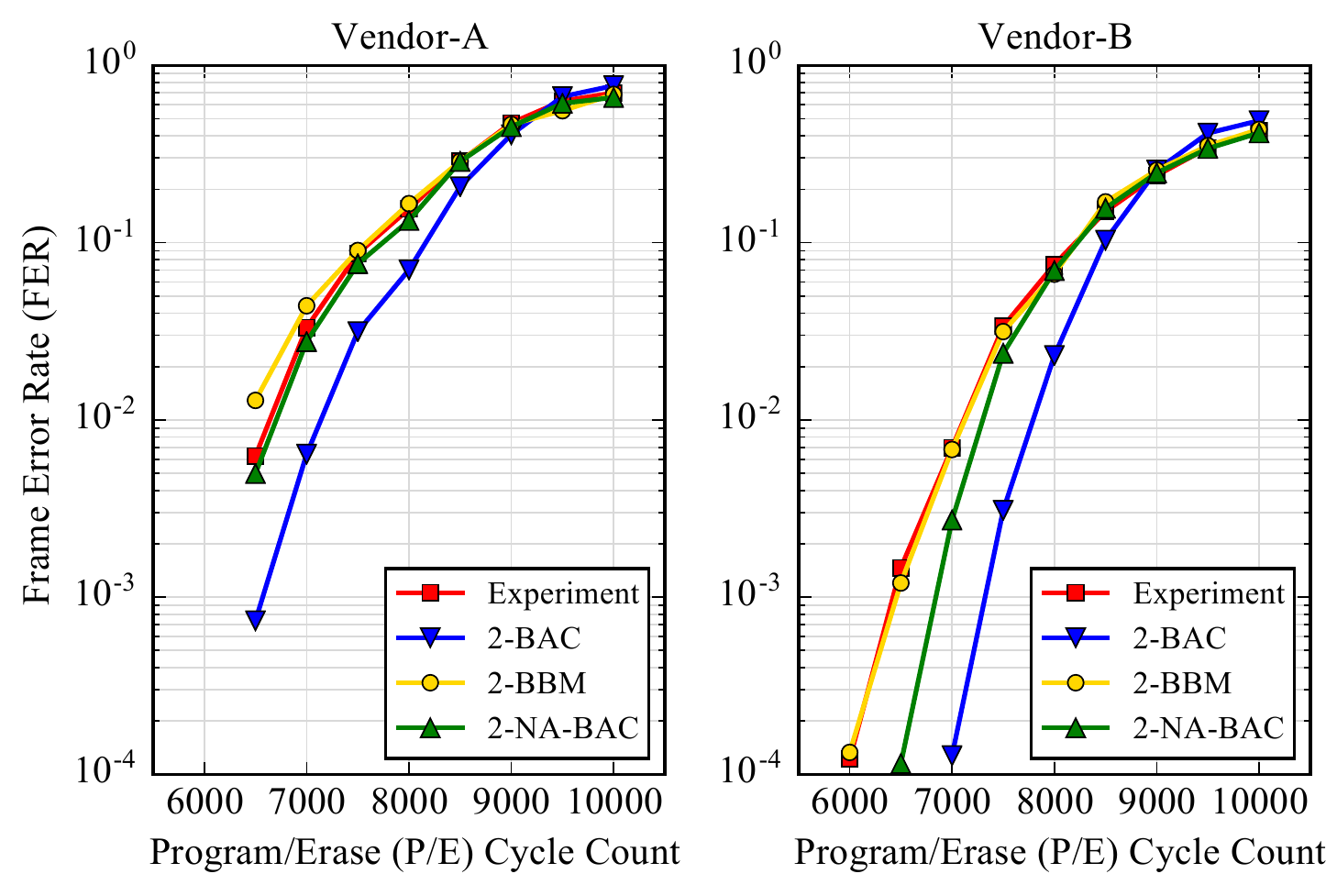}
		\caption{Comparison of FER performance of a ($N=8191$, $k=7683$, $t=39$) BCH code using empirical error data and error data from simulation using the 2-BAC, 2-BBM, 2-NA-BAC channel models for \mbox{vendor-A} and \mbox{vendor-B} chips.}
		\label{fig:bch_fer}
	\end{figure}
\else
	\begin{figure}[h]
		\centering
		\includegraphics[width=0.47\textwidth]{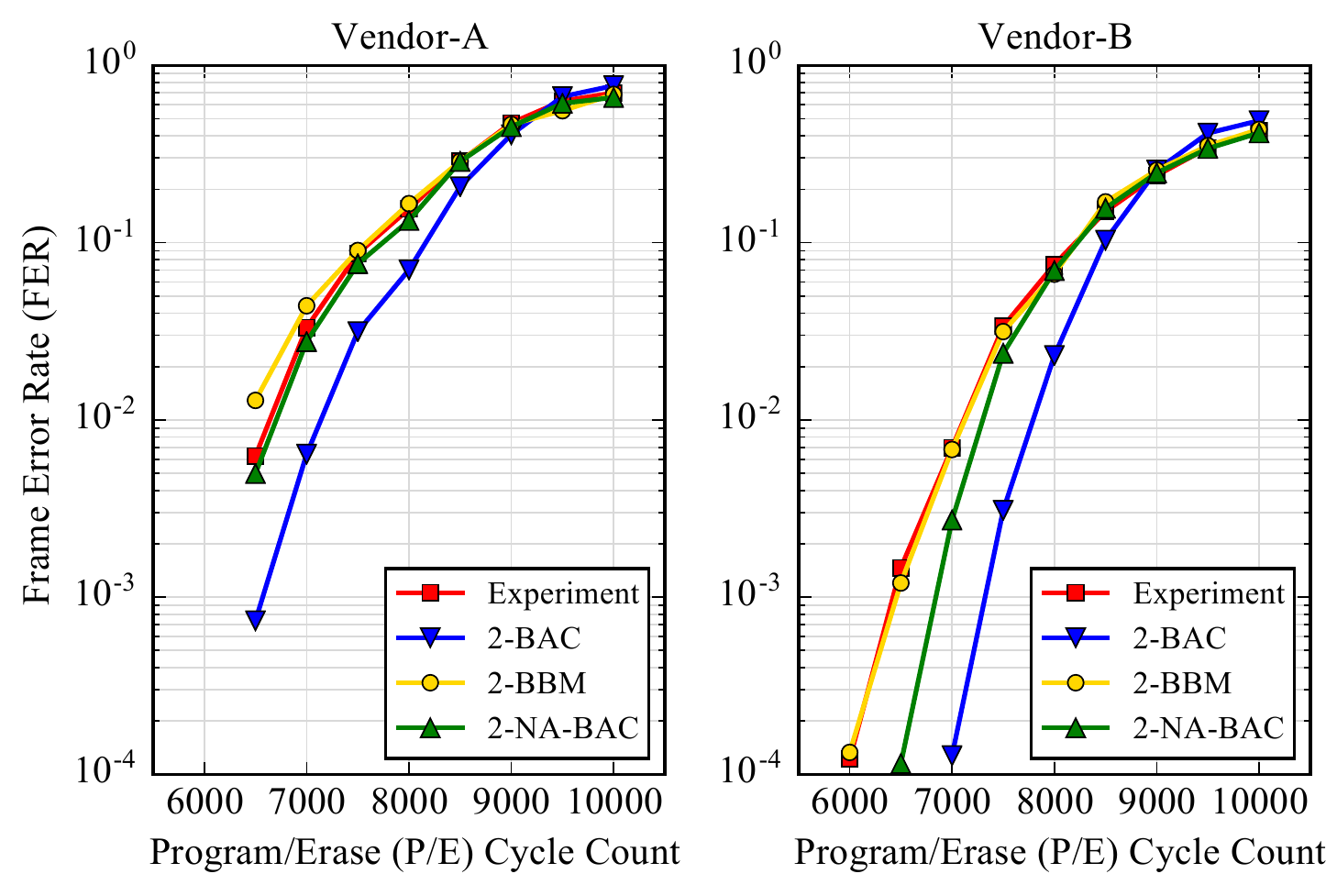}
		\caption{Comparison of FER performance of a ($N=8191$, $k=7683$, $t=39$) BCH code using empirical error data and error data from simulation using the 2-BAC, 2-BBM, 2-NA-BAC channel models for \mbox{vendor-A} and \mbox{vendor-B} chips.}
		\label{fig:bch_fer}
	\end{figure}
\fi

\ifCLASSOPTIONonecolumn
	\begin{figure}[h]
		\centering
		\includegraphics[width=0.7\textwidth]{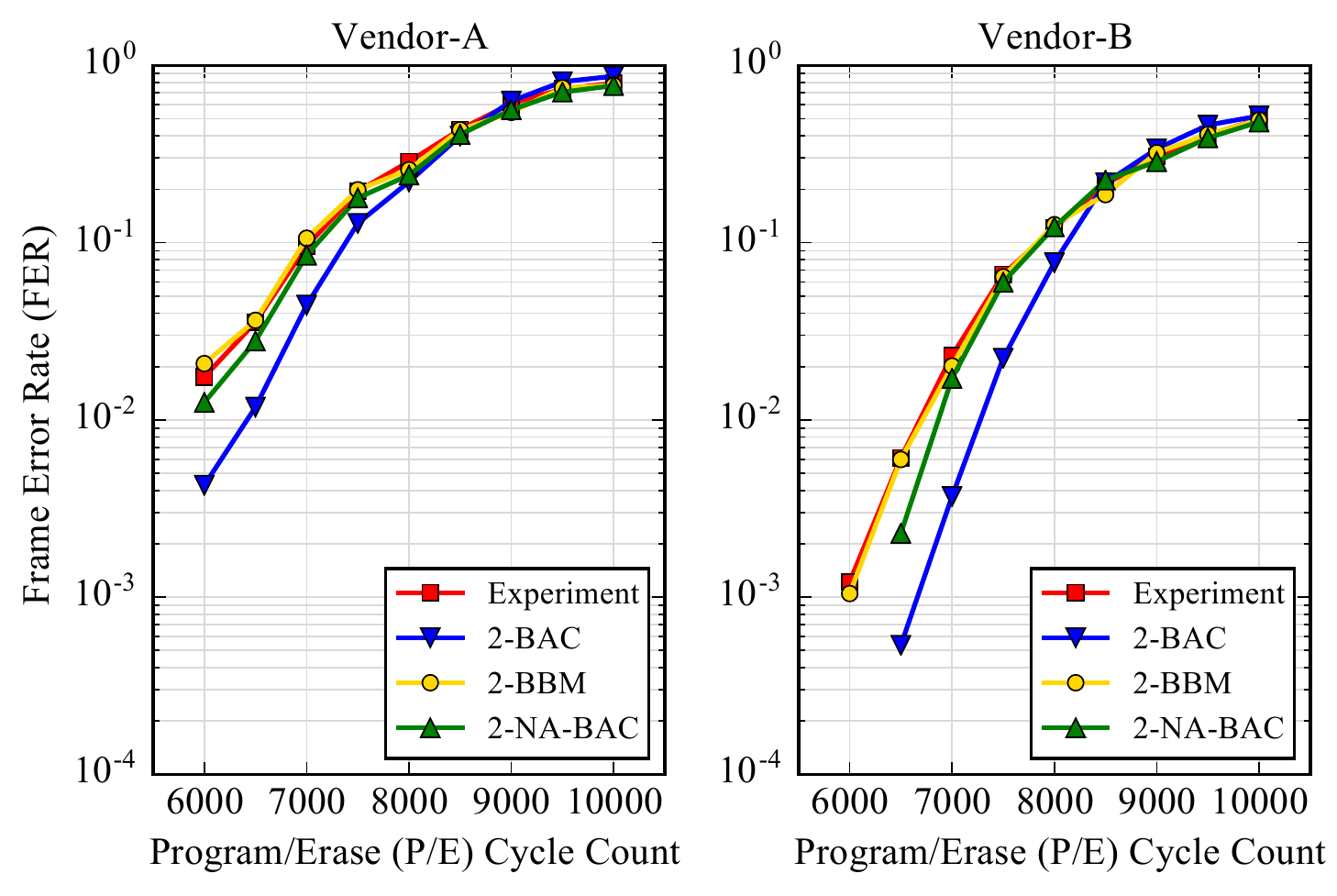}
		\caption{Comparison of FER performance of a ($N=8192$, $k=7683$) regular QC-LDPC code using empirical error data and error data from simulation using the 2-BAC, 2-BBM, 2-NA-BAC channel models for \mbox{vendor-A} and \mbox{vendor-B} chips.}
		\label{fig:ldpc_fer}
	\end{figure}
\else
	\begin{figure}[h]
		\centering
		\includegraphics[width=0.47\textwidth]{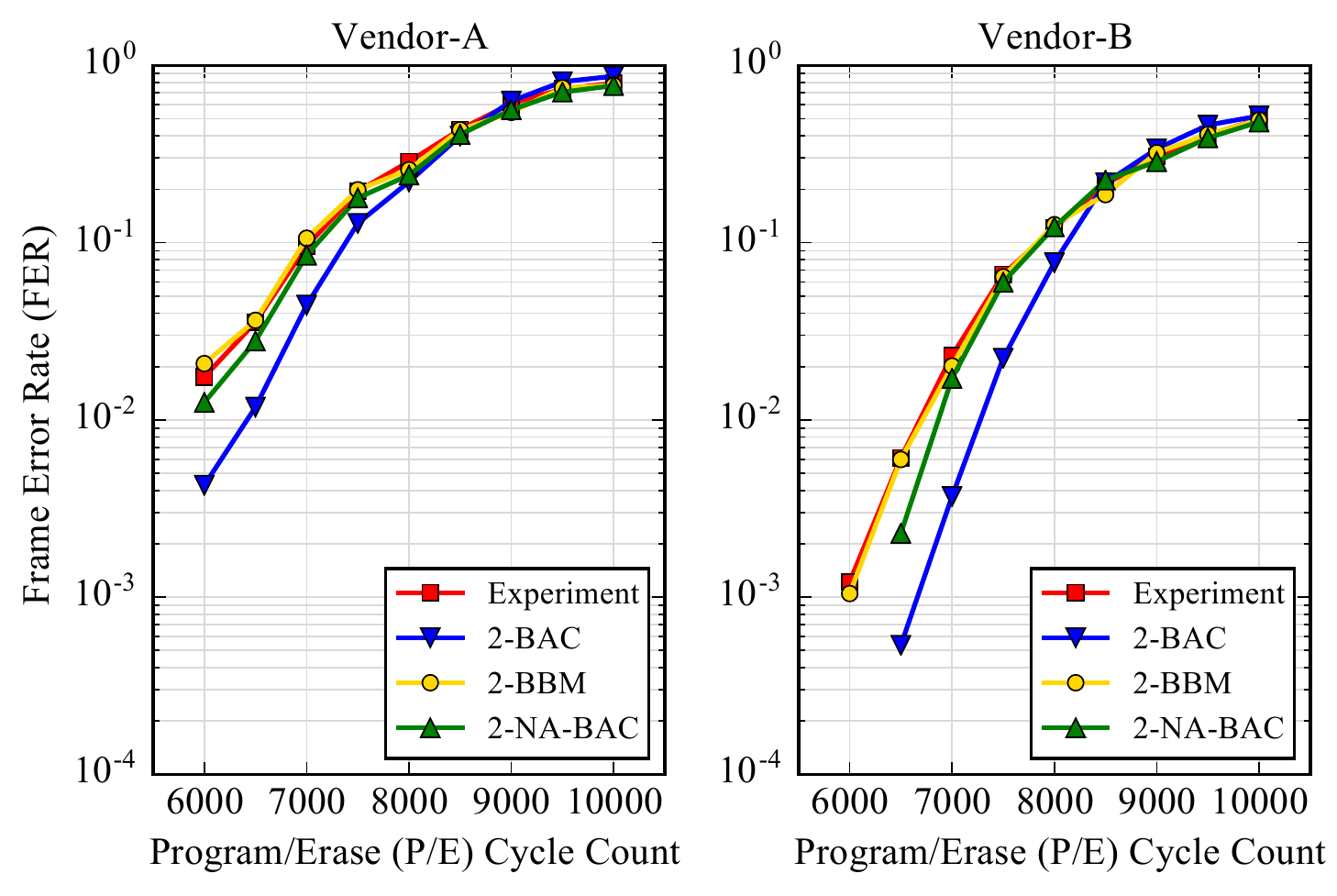}
		\caption{Comparison of FER performance of a ($N=8192$, $k=7683$) regular QC-LDPC code using empirical error data and error data from simulation using the 2-BAC, 2-BBM, 2-NA-BAC channel models for \mbox{vendor-A} and \mbox{vendor-B} chips.}
		\label{fig:ldpc_fer}
	\end{figure}
\fi

The FER performance of a ($N=8191$, $k=7683$, $t=39$) BCH code using empirical data and the proposed channel models is shown in Fig.~\ref{fig:bch_fer}. Fig.~\ref{fig:ldpc_fer} shows the FER performance of a ($N=8192$, $k=7683$) regular quasi-cyclic LDPC (QC-LDPC) code with $d_c = 64$ and $d_v = 4$, where $d_c$ and $d_v$ refer to the check node and variable node degrees, respectively, in the parity check matrix. The parity check matrix of the QC-LDPC code is constructed using size $128 \times 128$ circulant permutation matrices and the design rate is specified as $0.9375$. To ensure the required variable node degree $d_v$, exactly $d_v$ permutations of the circulant matrix are stacked vertically along the rows of the parity check matrix for every set of columns. Zero matrices of size $128 \times 128$
are used to fill up any remaining rows. This is done using the progressive edge growth (PEG) algorithm~\cite{Arnold_2001} to avoid short cycles. Note that although the specified design rate corresponds to a code dimension of $7680$, we get $k = 7683$ due to three dependent parity checks in the final parity check matrix thus obtained. A sum-product belief propagation decoder with a maximum of $50$ iterations and early termination is used to decode the QC-LDPC code. Fig.~\ref{fig:ldpc_fer_extra} also shows additional results comparing the FER performance of the QC-LDPC code obtained using empirical data and simulation data from the BAC, BBM channel models, separately for the lower and upper pages of \mbox{vendor-A} chip and the lower page of \mbox{vendor-B} chip. The lowest FER performance estimates from empirical data were obtained by P/E cycling $44$ and $24$ blocks of \mbox{vendor-A} and \mbox{vendor-B} chips, respectively. A total of $6$ and $4$ frame errors were observed to obtain the lowest FER performance estimates from empirical data for the lower and upper pages of \mbox{vendor-A} chip, respectively.
For the lower page of \mbox{vendor-B} chip, $4$ frame errors were observed to estimate the lowest FER performance from empirical data. Note that the results for the upper page of \mbox{vendor-B} chip are not shown as we did not observe any frame errors in the empirical data. We also note that a different \mbox{vendor-B} chip was used to obtain the additional results shown in Fig.~\ref{fig:ldpc_fer_extra} when compared to the rest of the paper.
Fig.~\ref{fig:polar_fer_sclist} shows the comparison of FER performance of a ($N=8192$, $k=7684$) polar code using empirical data and the proposed channel models. The polar code is optimized for a binary symmetric channel (BSC) with bit error probability $p=0.001$ using the construction technique proposed in~\cite{Tal_Vardy_2013}. The successive cancellation list (SC-List) decoder proposed in~\cite{Tal_Vardy_2015} is used for decoding the polar code.
\ifCLASSOPTIONonecolumn
	\begin{figure}
		\centering
		\includegraphics[width=0.7\textwidth]{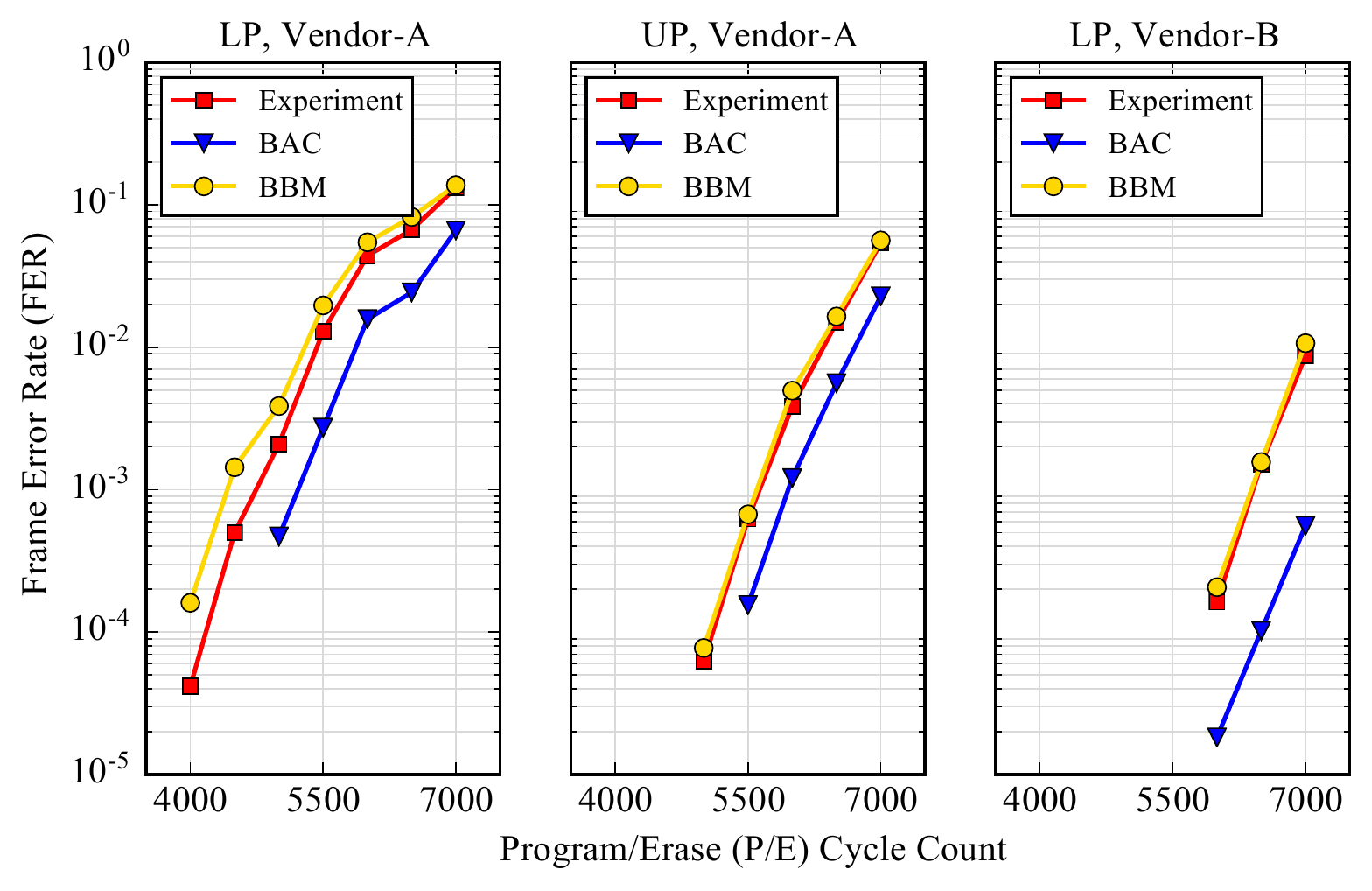}
		\caption{Comparison of FER performance of a ($N=8192$, $k=7683$) regular QC-LDPC code using empirical error data and error data from simulation using the BAC and BBM channel models for both lower and upper pages of \mbox{vendor-A} chip and the lower page of \mbox{vendor-B} chip.}
		\label{fig:ldpc_fer_extra}
	\end{figure}
\else
	\begin{figure}
		\centering
		\includegraphics[width=0.47\textwidth]{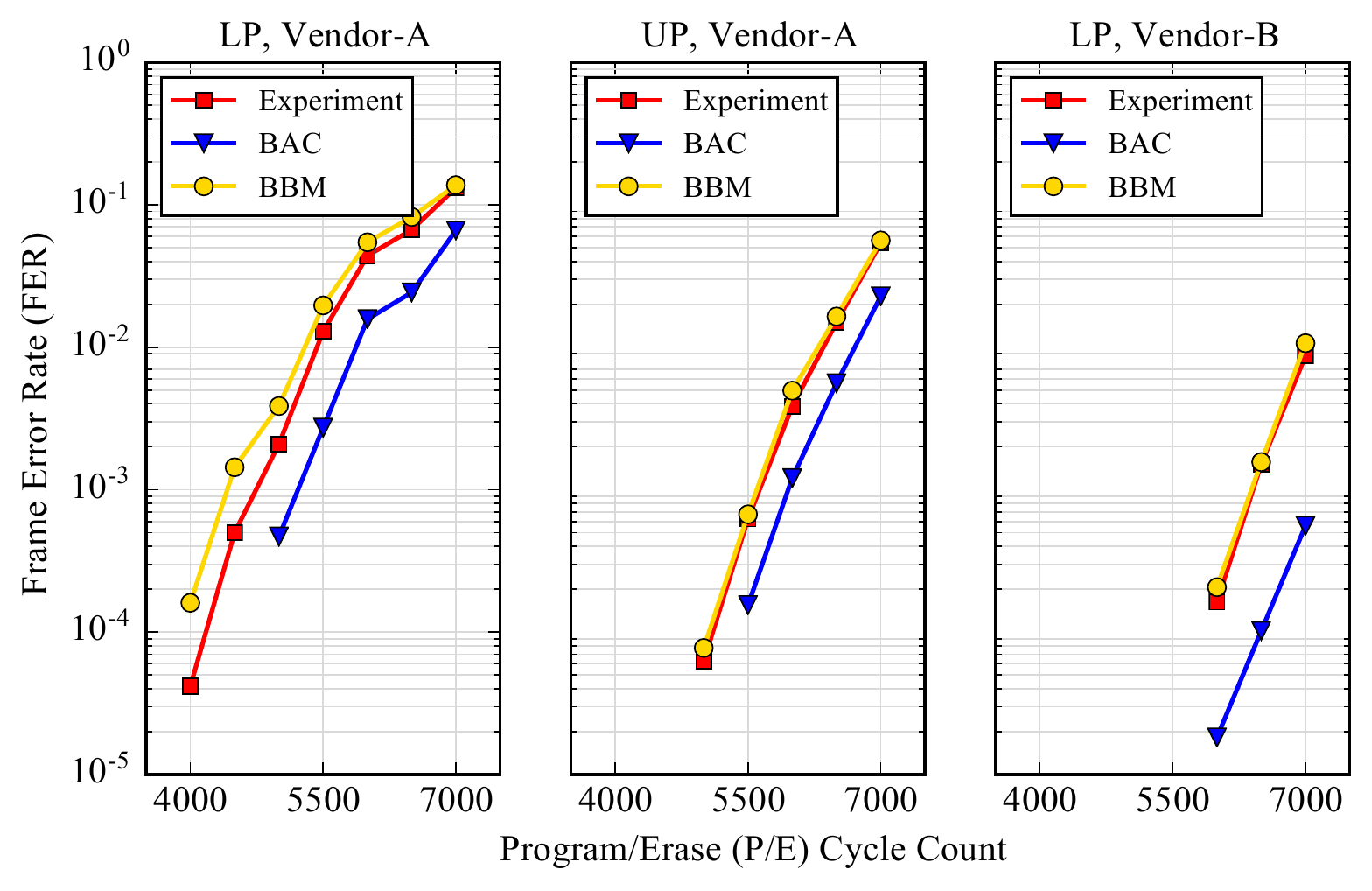}
		\caption{Comparison of FER performance of a ($N=8192$, $k=7683$) regular QC-LDPC code using empirical error data and error data from simulation using the BAC and BBM channel models for both lower and upper pages of \mbox{vendor-A} chip and the lower page of \mbox{vendor-B} chip.}
		\label{fig:ldpc_fer_extra}
	\end{figure}
\fi
\ifCLASSOPTIONonecolumn
	\begin{figure}
		\centering
		\includegraphics[width=0.7\textwidth]{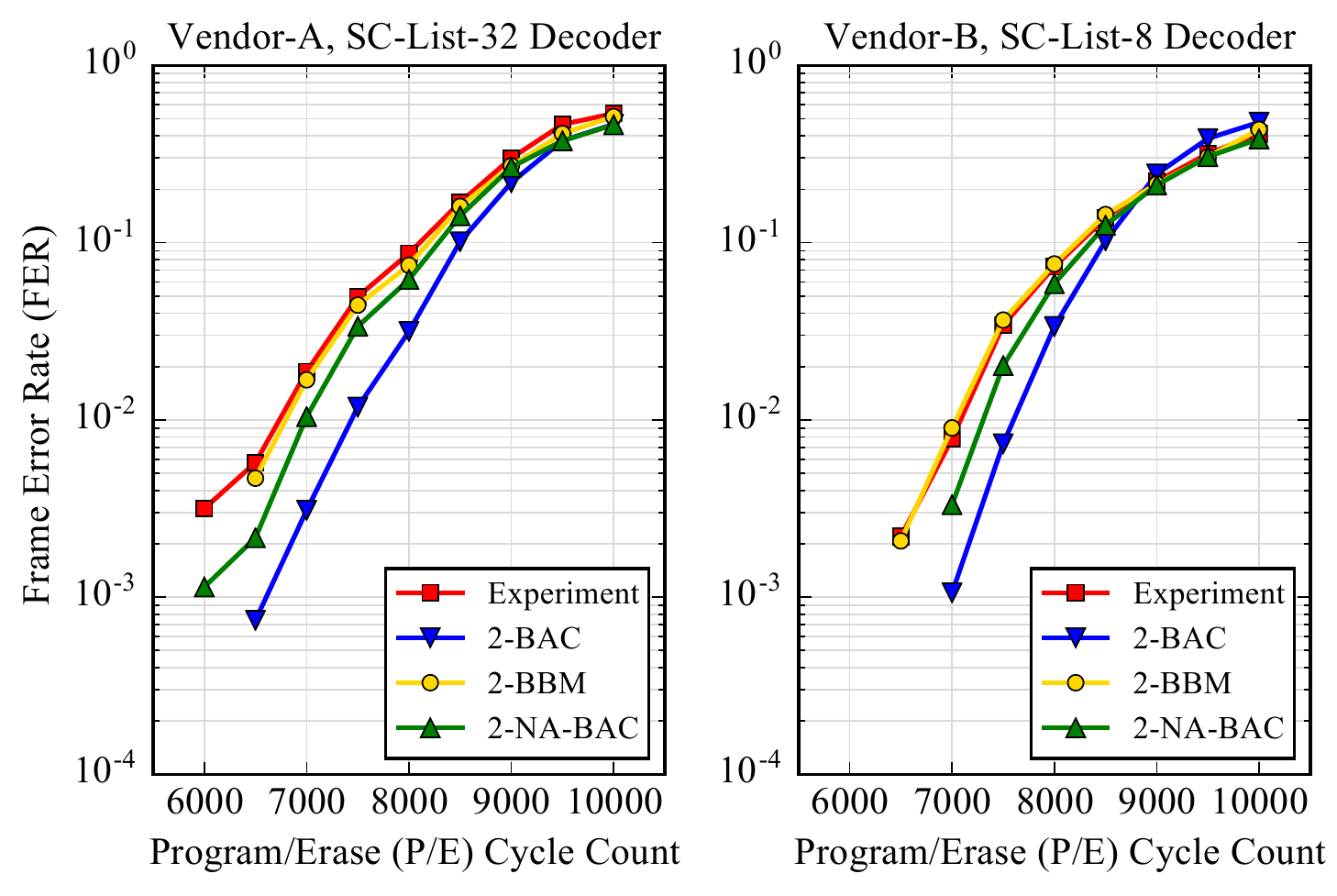}
		\caption{Comparison of FER performance of a ($N=8192$, $k=7684$) polar code optimized for BSC($0.001$) using empirical error data and error data from simulation using the 2-BAC, 2-BBM, 2-NA-BAC channel models for \mbox{vendor-A} and \mbox{vendor-B} chips. The SC-List decoder is used with a list size $= 32$ for \mbox{vendor-A} and list size $= 8$ for \mbox{vendor-B} chip.}
		\label{fig:polar_fer_sclist}
	\end{figure}
\else
	\begin{figure}
		\centering
		\includegraphics[width=0.47\textwidth]{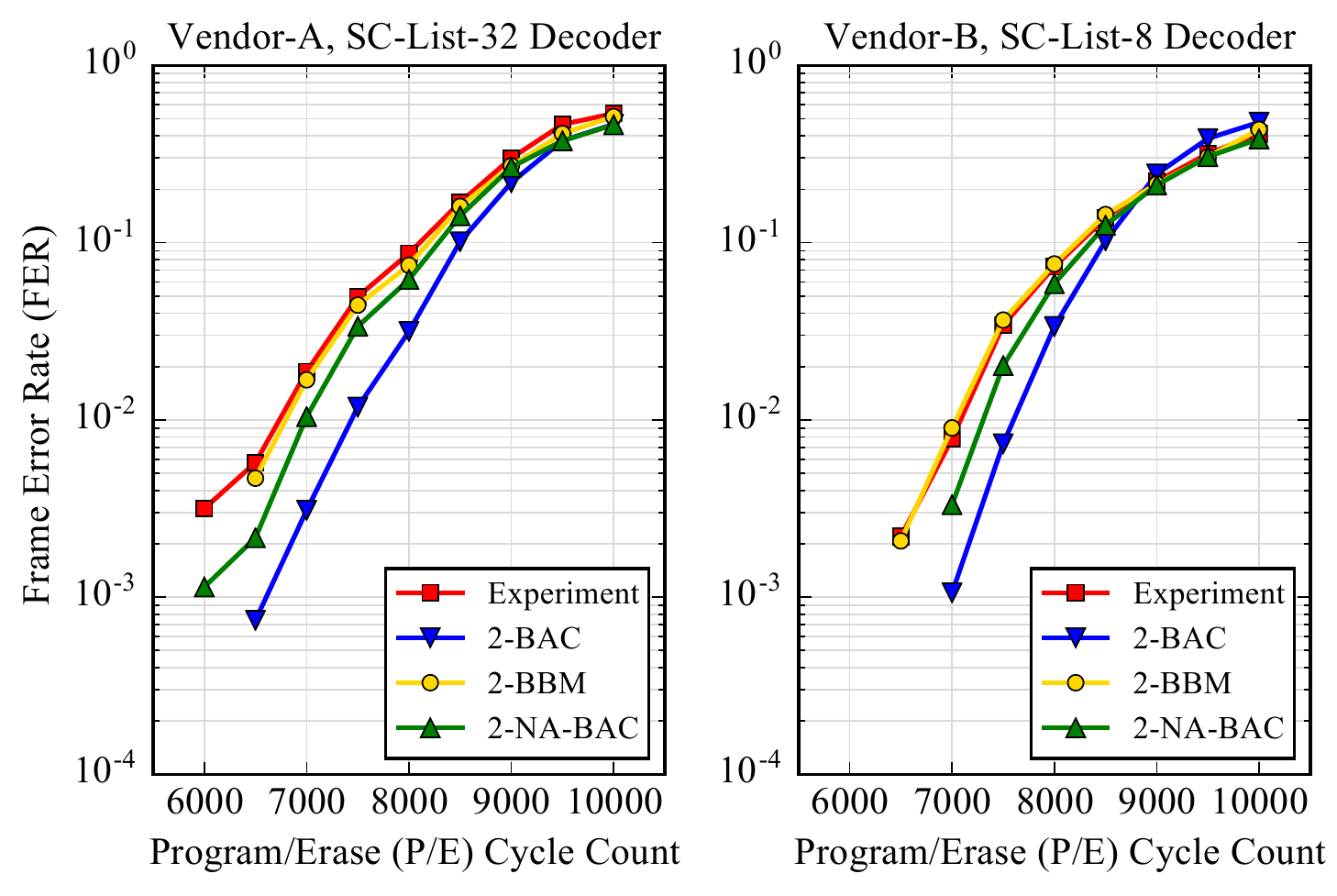}
		\caption{Comparison of FER performance of a ($N=8192$, $k=7684$) polar code optimized for BSC($0.001$) using empirical error data and error data from simulation using the 2-BAC, 2-BBM, 2-NA-BAC channel models for \mbox{vendor-A} and \mbox{vendor-B} chips. The SC-List decoder is used with a list size $= 32$ for \mbox{vendor-A} and list size $= 8$ for \mbox{vendor-B} chip.}
		\label{fig:polar_fer_sclist}
	\end{figure}
\fi

For all the ECCs considered and using data from both vendor chips, we observe that the 2-BAC model provides an optimistic estimate of the FER performance when compared to the empirically observed FER performance. This is mainly due to the inability of the \mbox{2-BAC} model to capture the high variance in the number of bit errors per frame observed empirically. The gap in ECC FER performance estimates using the 2-BAC model and the empirical data is increasing as the FER decreases, and it is about an order of magnitude for \mbox{vendor-A} chip at $6,500$ P/E cycles and greater than an order of magnitude for \mbox{vendor-B} chip at $7,000$ P/E cycles for the BCH code as shown in Fig.~\ref{fig:bch_fer}. This gap in ECC FER performance estimates at low FERs is bad for determining the correct endurance (life-time) of a flash memory chip. From the results shown in~Fig.~\ref{fig:ldpc_fer_extra} for the QC-LDPC code, we observe that the BBM channel model estimates the FER performance accurately even at lower FERs around $10^{-4}$, for the upper page of \mbox{vendor-A} chip and the lower page of \mbox{vendor-B} chip. The FER performance estimates obtained using the BBM channel model are better than those obtained using the BAC channel model for the lower page of \mbox{vendor-A} chip, however we observe a small mismatch in the BBM channel model FER performance estimates at lower FERs when compared to the empirical FER estimates. This mismatch is due to the inability of the BBM channel model to fit the larger proportions of frames with small number of bit errors observed in the lower tail of the empirical error histograms for the lower page of \mbox{vendor-A} chip.
This appears to be a vendor-specific effect, as this kind of effect was not observed in the empirical error histograms corresponding to the lower page of \mbox{vendor-B} chip. Overall, the \mbox{2-BBM} model is able to match the empirical ECC FER performance estimates accurately, while the estimates obtained using the \mbox{2-NA-BAC} model lie between those of the \mbox{2-BAC} and the \mbox{2-BBM} models. The ECC FER performance estimates using the \mbox{2-PA-BAC} model are the same as those using the \mbox{2-NA-BAC} model and are omitted. From these results it is clear that the proposed \mbox{2-BBM} channel model is able to accurately describe the nature of the number of bit errors per frame in MLC flash memories and hence provides accurate estimates of the ECC FER performance.

%% file: sections/conclusion.tex
\section{Conclusion}
\label{sec:conclusion}

We studied the feasibility of using well known discrete memoryless channel models to model the MLC flash memory channel. Based on empirical error analysis and ECC FER performance estimation for BCH, LDPC, and polar codes, we observe that the 2-BAC model with parameter estimates derived from empirical error data suffices to produce an accurate estimate of the average raw bit error rate, but it provides an incorrect optimistic estimate of the ECC FER performance when compared to the empirically observed ECC FER performance.
This is mainly due to the overdispersed nature of the number of bit errors per frame in empirical data which is not modeled well by the 2-BAC model. We proposed the 2-Beta-Binomial (2-BBM) channel model based on the beta-binomial probability distribution and using statistical analysis, goodness of fit tests and ECC FER performance results showed that the 2-BBM channel model accurately describes the nature of the number of bit errors per frame in MLC flash memories. We also note that the BBM channel model can be shown to be equivalent to an urn based channel model~\cite{Alajaji_1994} and hence has memory associated with it. Although not presented in this paper, our preliminary experiment results for combined data retention plus P/E cycling stress show the evidence of overdispersion in error statistics and the suitability of the proposed 2-BBM channel model. We leave a detailed examination of this as future work.
Although the proposed channel models are for MLC flash memories, the proposed empirical design approach is generic and can easily be extended for three-level cell (TLC) flash memories.

%% file: sections/appendix_singlecol.tex
\appendices
\section{Proof of Proposition \ref{prop:bac_k_mean_var}}
\label{app:proof_bac_k_mean_var}
To compute $\Var[K]$, we compute its mean $\E[K]$ and the second moment $\E[K^2]$. Based on (\ref{eqn:k_dist}), both these moments of $K$ can be computed from the moments of $\kcountm{m}$ as
\begin{IEEEeqnarray}{rCl}
   \E[K] & = & \sum_{m=0}^{N}\frac{{N \choose m}}{2^{N}}~\E[\kcountm{m}] \\
   \E[K^2] & = & \sum_{m=0}^{N}\frac{{N \choose m}}{2^{N}}~\E[\kcountm{m}^2].
\end{IEEEeqnarray}
From (\ref{eqn:km_sum}) and (\ref{eqn:xy_indep}), we have
\begin{IEEEeqnarray}{rCl}
   \E[\kcountm{m}] & = & \E[\kcount{m}{0}] + \E[\kcount{N-m}{1}] \nonumber \\
				   & = & mp + (N-m)q \label{eqn:km_mean} \\
   \Var[\kcountm{m}] & = & \Var[\kcount{m}{0}] + \Var[\kcount{N-m}{1}] \nonumber \\
					& = & mp(1-p) + (N-m)q(1-q). \label{eqn:km_var}
\end{IEEEeqnarray}
Therefore, $\E[\kcountm{m}^2]$ is given by
\begin{IEEEeqnarray}{rCl}
   \E[\kcountm{m}^2] & = & \Var[\kcountm{m}] + (\E[\kcountm{m}])^2 \nonumber \\
             & = & mp + (N-m)q + m(m-1)p^2 + 2m(N-m)pq + (N-m)(N-m-1)q^2.
   \label{eqn:zm_moment_2}
\end{IEEEeqnarray}
Hence $\E[K]$ and $\E[K^2]$ are given by
\begin{IEEEeqnarray}{rCl}
   \E[K] & = & \sum_{m=0}^{N}\frac{{N \choose m}}{2^{N}}~\E[\kcountm{m}] \nonumber \\
         & = & \frac{N}{2}(p + q)
   \label{eqn:k_mean}\\
   \E[K^2] & = & \sum_{m=0}^{N}\frac{{N \choose m}}{2^{N}}~\E[\kcountm{m}^2] \nonumber \\
         & = & \frac{N}{2}(p + q) + \Big(\frac{N^2 - N}{2}\Big)pq + \Big(\frac{N^2 - N}{4}\Big)(p^2 + q^2).
   \label{eqn:k_moment_2}
\end{IEEEeqnarray}
Note that we have used the combinatorial identities
\begin{IEEEeqnarray}{L}
   \sum_{m=0}^{N}{N \choose m}~m = N 2^{N-1} \label{eqn:comb_id_1} \\
   \sum_{m=0}^{N}{N \choose m}~m^2 = (N + N^2) 2^{N-2}. \label{eqn:comb_id_2}
\end{IEEEeqnarray}
Therefore we can obtain $\Var[K]$ from (\ref{eqn:k_mean}) and (\ref{eqn:k_moment_2}) as
\begin{IEEEeqnarray}{rCl+x*}
   \Var[K] & = & \E[K^2] - (\E[K])^2 \nonumber \\
           & = & \frac{N}{2}\Big((p + q) - pq - \frac{1}{2}(p^2 + q^2) \Big). \\* &&& \IEEEQED \nonumber
\end{IEEEeqnarray}
\section{Proof of Proposition \ref{prop:bbm_k_mean_var}}
\label{app:proof_bbm_k_mean_var}
We take the same approach as the proof of Proposition~1.
From (\ref{eqn:km_sum}) and (\ref{eqn:bbm_xy_indep}), we have
\begin{IEEEeqnarray}{rCl}
   \E[\kcountm{m}] & = & \Bigg(\frac{ma}{a+b}\Bigg) + \Bigg(\frac{(N-m)c}{c+d}\Bigg) \label{eqn:bbm_km_mean} \\
   \Var[\kcountm{m}] & = & \Bigg(\frac{mab(a+b+m)}{(a+b)^2(a+b+1)}\Bigg) + \Bigg(\frac{(N-m)cd(c+d+N-m)}{(c+d)^2(c+d+1)}\Bigg). \label{eqn:bbm_km_var}
\end{IEEEeqnarray}
Therefore, $\E[\kcountm{m}^2]$ is given by
\begin{IEEEeqnarray}{rCl}
    \E[\kcountm{m}^2] & = & \Var[\kcountm{m}] + (\E[\kcountm{m}])^2 \nonumber \\
                      & = & \Var[\kcount{m}{0}] + (\E[\kcount{m}{0}])^2 + \Var[\kcount{N-m}{1}] + (\E[\kcount{N-m}{1}])^2 + 2\E[\kcount{m}{0}]\E[\kcount{N-m}{1}].
    \label{eqn:bbm_km_moment_2}
\end{IEEEeqnarray}
Substituting using (\ref{eqn:xm_bbm_mean}) - (\ref{eqn:ynm_bbm_var}) and simplifying, we have
\begin{IEEEeqnarray}{rCl}
    \E[\kcountm{m}^2] & = & \Bigg(\frac{ma(m(a+1) + b)}{(a+b)(a+b+1)}\Bigg) + \Bigg(\frac{(N-m)c((N-m)(c+1) + d)}{(c+d)(c+d+1)}\Bigg) \nonumber \\
    &   & + \Bigg(\frac{2m(N-m)ac}{(a+b)(c+d)}\Bigg).
\end{IEEEeqnarray}
Hence $\E[K]$ and $\E[K^2]$ are given by
\begin{IEEEeqnarray}{rCl}
   \E[K] & = & \sum_{m=0}^{N}\frac{{N \choose m}}{2^{N}}~\E[\kcountm{m}] \nonumber \\
         & = & \frac{N}{2}\Bigg(\frac{a}{a+b} + \frac{c}{c+d} \Bigg)
    \label{eqn:bbm_k_mean} \\
   \E[K^2] & = & \sum_{m=0}^{N}\frac{{N \choose m}}{2^{N}}~\E[\kcountm{m}^2] \nonumber \\
         & = & \frac{N}{4}\Bigg(\frac{(N+1)a(a+1) + 2Nab}{(a+b)(a+b+1)}\Bigg) + \frac{N}{4}\Bigg(\frac{(N+1)c(c+1) + 2Ncd}{(c+d)(c+d+1)}\Bigg) \nonumber \\
         &   & + \frac{N(N-1)}{4}\Bigg(\frac{2ac}{(a+b)(c+d)}\Bigg).
   \label{eqn:bbm_k_moment_2}
\end{IEEEeqnarray}
We have used the combinatorial identities (\ref{eqn:comb_id_1}) and (\ref{eqn:comb_id_2}). From (\ref{eqn:bbm_k_mean}) and (\ref{eqn:bbm_k_moment_2}), $\Var[K]$ is easily obtained as
\begin{IEEEeqnarray}{rCl+x*}
   \Var[K] & = & \E[K^2] - (\E[K])^2 \nonumber \\
           & = & \frac{N}{4}\Bigg(\frac{a(a+b)(a+2b+1) + Nab}{(a+b)^2(a+b+1)}\Bigg) + \frac{N}{4}\Bigg(\frac{c(c+d)(c+2d+1) + Ncd}{(c+d)^2(c+d+1)}\Bigg) \nonumber \\
           &  & - \frac{N}{4}\Bigg(\frac{2ac}{(a+b)(c+d)} \Bigg). \\* &&& \IEEEQED \nonumber
\end{IEEEeqnarray}

\section{Proof of Proposition \ref{prop:bbm_k0_k1_mean_second_moment}}
\label{app:proof_bbm_k0_k1_mean_second_moment}
This proof proceeds along similar lines as the proof of Proposition~2.
From (\ref{eqn:k0_dist}) and (\ref{eqn:bbm_km0_dist}),
\begin{IEEEeqnarray}{rCl}
    \Pr(K^{(0)} = k) & = & \sum_{m=k}^{N}\frac{{N \choose m}}{2^N}{m \choose k}\frac{B(a+k, b+m-k)}{B(a, b)} \nonumber
\end{IEEEeqnarray}
\begin{IEEEeqnarray}{rCl}
    \E[K^{(0)}] & = & \sum_{k = 0}^{N} k \Pr(K^{(0)} = k) \nonumber \\
                & = & \frac{1}{2^N} \sum_{k = 0}^{N} \sum_{m=k}^{N} k {N \choose m}{m \choose k}\frac{B(a+k, b+m-k)}{B(a, b)}  \nonumber \\
                & = & \frac{1}{2^N} \sum_{m = 0}^{N} {N \choose m} \sum_{k = 0}^{m} k {m \choose k}\frac{B(a+k, b+m-k)}{B(a, b)}  \nonumber \\
                & = & \frac{1}{2^N} \sum_{m = 0}^{N} {N \choose m} \E[\kcount{m}{0}] \nonumber \\
                & = & \frac{N}{2}\Bigg(\frac{a}{a+b}\Bigg) \\
    \E[(K^{(0)})^2] & = & \sum_{k = 0}^{N} k^2 \Pr(K^{(0)} = k) \nonumber \\
                & = & \frac{1}{2^N} \sum_{k = 0}^{N} \sum_{m=k}^{N} k^2 {N \choose m}{m \choose k}\frac{B(a+k, b+m-k)}{B(a, b)}  \nonumber \\
                & = & \frac{1}{2^N} \sum_{m = 0}^{N} {N \choose m} \sum_{k = 0}^{m} k^2 {m \choose k}\frac{B(a+k, b+m-k)}{B(a, b)}  \nonumber \\
                & = & \frac{1}{2^N} \sum_{m = 0}^{N} {N \choose m} \E[(\kcount{m}{0})^2] \nonumber \\
                & = & \frac{N}{4}\Bigg(\frac{a(a+2b+1) + Na(a+1)}{(a+b)(a+b+1)}\Bigg) \\
    \Var[K^{(0)}] & = & \E[(K^{(0)})^2] - (\E[K^{(0)}])^2 \nonumber \\
                  & = & \frac{N}{4}\Bigg(\frac{a(a+b)(a+2b+1) + Nab}{(a+b)^2(a+b+1)}\Bigg).
\end{IEEEeqnarray}
We have used the combinatorial identities (\ref{eqn:comb_id_1}) and (\ref{eqn:comb_id_2}) and also the fact that the second moment of a beta-binomial random variable $\kcount{m}{0} \sim \textrm{Beta-Binomial}(m, a, b)$ is given by $\frac{ma(m(a+1) + b)}{(a+b)(a+b+1)}$. The expressions for $\E[K^{(1)}]$ and $\Var[K^{(1)}]$ can be derived similarly.~\hfill\IEEEQED

%% file: sections/appendix.tex
\appendices
\section{Proof of Proposition \ref{prop:bac_k_mean_var}}
\label{app:proof_bac_k_mean_var}
To compute $\Var[K]$, we compute its mean $\E[K]$ and the second moment $\E[K^2]$. Based on (\ref{eqn:k_dist}), both these moments of $K$ can be computed from the moments of $\kcountm{m}$ as
\begin{IEEEeqnarray}{C}
   \E[K] = \sum_{m=0}^{N}\frac{{N \choose m}}{2^{N}}~\E[\kcountm{m}] \\
   \E[K^2] = \sum_{m=0}^{N}\frac{{N \choose m}}{2^{N}}~\E[\kcountm{m}^2].
\end{IEEEeqnarray}
From (\ref{eqn:km_sum}) and (\ref{eqn:xy_indep}), we have
\begin{IEEEeqnarray}{rCl}
   \E[\kcountm{m}] & = & \E[\kcount{m}{0}] + \E[\kcount{N-m}{1}] \nonumber \\
				   & = & mp + (N-m)q \label{eqn:km_mean} \\
   \Var[\kcountm{m}] & = & \Var[\kcount{m}{0}] + \Var[\kcount{N-m}{1}] \nonumber \\
					& = & mp(1-p) + (N-m)q(1-q). \label{eqn:km_var}
\end{IEEEeqnarray}
Therefore, $\E[\kcountm{m}^2]$ is given by
\begin{IEEEeqnarray}{rCl}
   \E[\kcountm{m}^2] & = & \Var[\kcountm{m}] + (\E[\kcountm{m}])^2 \nonumber \\
             & = & mp + (N-m)q + m(m-1)p^2 + 2m(N-m)pq\nonumber \\
             &   & + (N-m)(N-m-1)q^2.
   \label{eqn:zm_moment_2}
\end{IEEEeqnarray}
Hence $\E[K]$ and $\E[K^2]$ are given by
\begin{IEEEeqnarray}{rCl}
   \E[K] & = & \sum_{m=0}^{N}\frac{{N \choose m}}{2^{N}}~\E[\kcountm{m}] \nonumber \\
         & = & \frac{N}{2}(p + q)
   \label{eqn:k_mean} \\
   \E[K^2] & = & \sum_{m=0}^{N}\frac{{N \choose m}}{2^{N}}~\E[\kcountm{m}^2] \nonumber \\
         & = & \frac{N}{2}(p + q) + \Big(\frac{N^2 - N}{2}\Big)pq \nonumber \\
         &   & + \Big(\frac{N^2 - N}{4}\Big)(p^2 + q^2). \label{eqn:k_moment_2}
\end{IEEEeqnarray}
Note that we have used the combinatorial identities
\begin{IEEEeqnarray}{L}
   \sum_{m=0}^{N}{N \choose m}~m = N 2^{N-1} \label{eqn:comb_id_1} \\
   \sum_{m=0}^{N}{N \choose m}~m^2 = (N + N^2) 2^{N-2}. \label{eqn:comb_id_2}
\end{IEEEeqnarray}
Therefore we can obtain $\Var[K]$ from (\ref{eqn:k_mean}) and (\ref{eqn:k_moment_2}) as
\begin{IEEEeqnarray}{rCl+x*}
   \Var[K] & = & \E[K^2] - (\E[K])^2 \nonumber \\
           & = & \frac{N}{2}\Big((p + q) - pq - \frac{1}{2}(p^2 + q^2) \Big). \\* &&& \IEEEQED \nonumber
\end{IEEEeqnarray}
\section{Proof of Proposition \ref{prop:bbm_k_mean_var}}
\label{app:proof_bbm_k_mean_var}
We take the same approach as the proof of Proposition~1.
From (\ref{eqn:km_sum}) and (\ref{eqn:bbm_xy_indep}), we have
\begin{IEEEeqnarray}{rCl}
   \E[\kcountm{m}] & = & \Bigg(\frac{ma}{a+b}\Bigg) + \Bigg(\frac{(N-m)c}{c+d}\Bigg) \label{eqn:bbm_km_mean} \\
   \Var[\kcountm{m}] & = & \Bigg(\frac{mab(a+b+m)}{(a+b)^2(a+b+1)}\Bigg) + \nonumber \\
                    &   & \Bigg(\frac{(N-m)cd(c+d+N-m)}{(c+d)^2(c+d+1)}\Bigg). \label{eqn:bbm_km_var}
\end{IEEEeqnarray}
Therefore, $\E[\kcountm{m}^2]$ is given by
\begin{IEEEeqnarray}{rCl}
    \E[\kcountm{m}^2] & = & \Var[\kcountm{m}] + (\E[\kcountm{m}])^2 \nonumber \\
                      & = & \Var[\kcount{m}{0}] + (\E[\kcount{m}{0}])^2 + \Var[\kcount{N-m}{1}] + \nonumber \\
                      &   & (\E[\kcount{N-m}{1}])^2 + 2\E[\kcount{m}{0}]\E[\kcount{N-m}{1}].
    \label{eqn:bbm_km_moment_2}
\end{IEEEeqnarray}
Substituting using (\ref{eqn:xm_bbm_mean}) - (\ref{eqn:ynm_bbm_var}) and simplifying, we have
\begin{IEEEeqnarray}{rCl}
    \E[\kcountm{m}^2] & = & \Bigg(\frac{ma(m(a+1) + b)}{(a+b)(a+b+1)}\Bigg) + \nonumber \\
                      &   & \Bigg(\frac{(N-m)c((N-m)(c+1) + d)}{(c+d)(c+d+1)}\Bigg) + \nonumber \\
                      &   & \Bigg(\frac{2m(N-m)ac}{(a+b)(c+d)}\Bigg).
\end{IEEEeqnarray}
Hence $\E[K]$ and $\E[K^2]$ are given by
\begin{IEEEeqnarray}{rCl}
   \E[K] & = & \sum_{m=0}^{N}\frac{{N \choose m}}{2^{N}}~\E[\kcountm{m}] \nonumber \\
         & = & \frac{N}{2}\Bigg(\frac{a}{a+b} + \frac{c}{c+d} \Bigg)
    \label{eqn:bbm_k_mean} \\
   \E[K^2] & = & \sum_{m=0}^{N}\frac{{N \choose m}}{2^{N}}~\E[\kcountm{m}^2] \nonumber \\
         & = & \frac{N}{4}\Bigg(\frac{(N+1)a(a+1) + 2Nab}{(a+b)(a+b+1)}\Bigg) + \nonumber \\
         &   & \frac{N}{4}\Bigg(\frac{(N+1)c(c+1) + 2Ncd}{(c+d)(c+d+1)}\Bigg) + \nonumber \\
         &   & \frac{N(N-1)}{4}\Bigg(\frac{2ac}{(a+b)(c+d)}\Bigg).
   \label{eqn:bbm_k_moment_2}
\end{IEEEeqnarray}
We have used the combinatorial identities (\ref{eqn:comb_id_1}) and (\ref{eqn:comb_id_2}). From (\ref{eqn:bbm_k_mean}) and (\ref{eqn:bbm_k_moment_2}), $\Var[K]$ is easily obtained as
\begin{IEEEeqnarray}{rCl+x*}
   \Var[K] & = & \E[K^2] - (\E[K])^2 \nonumber \\
           & = & \frac{N}{4}\Bigg(\frac{a(a+b)(a+2b+1) + Nab}{(a+b)^2(a+b+1)}\Bigg) + \nonumber \\
           &   & \frac{N}{4}\Bigg(\frac{c(c+d)(c+2d+1) + Ncd}{(c+d)^2(c+d+1)}\Bigg) - \nonumber \\
   		   &   & \frac{N}{4}\Bigg(\frac{2ac}{(a+b)(c+d)} \Bigg). \\* &&& \IEEEQED \nonumber
\end{IEEEeqnarray}

\section{Proof of Proposition \ref{prop:bbm_k0_k1_mean_second_moment}}
\label{app:proof_bbm_k0_k1_mean_second_moment}
This proof proceeds along similar lines as the proof of Proposition~2.
From (\ref{eqn:k0_dist}) and (\ref{eqn:bbm_km0_dist}),
\begin{IEEEeqnarray}{rCl}
    \Pr(K^{(0)} = k) & = & \sum_{m=k}^{N}\frac{{N \choose m}}{2^N}{m \choose k}\frac{B(a+k, b+m-k)}{B(a, b)} \nonumber
\end{IEEEeqnarray}
\begin{IEEEeqnarray}{rCl}
    \E[K^{(0)}] & = & \sum_{k = 0}^{N} k \Pr(K^{(0)} = k) \nonumber \\
                & = & \frac{1}{2^N} \sum_{k = 0}^{N} \sum_{m=k}^{N} k {N \choose m}{m \choose k}\frac{B(a+k, b+m-k)}{B(a, b)}  \nonumber \\
                & = & \frac{1}{2^N} \sum_{m = 0}^{N} {N \choose m} \sum_{k = 0}^{m} k {m \choose k}\frac{B(a+k, b+m-k)}{B(a, b)}  \nonumber \\
                & = & \frac{1}{2^N} \sum_{m = 0}^{N} {N \choose m} \E[\kcount{m}{0}] \nonumber \\
                & = & \frac{N}{2}\Bigg(\frac{a}{a+b}\Bigg) \\
    \E[(K^{(0)})^2] & = & \sum_{k = 0}^{N} k^2 \Pr(K^{(0)} = k) \nonumber \\
                & = & \frac{1}{2^N} \sum_{k = 0}^{N} \sum_{m=k}^{N} k^2 {N \choose m}{m \choose k}\frac{B(a+k, b+m-k)}{B(a, b)}  \nonumber \\
                & = & \frac{1}{2^N} \sum_{m = 0}^{N} {N \choose m} \sum_{k = 0}^{m} k^2 {m \choose k}\frac{B(a+k, b+m-k)}{B(a, b)}  \nonumber \\
                & = & \frac{1}{2^N} \sum_{m = 0}^{N} {N \choose m} \E[(\kcount{m}{0})^2] \nonumber \\
                & = & \frac{N}{4}\Bigg(\frac{a(a+2b+1) + Na(a+1)}{(a+b)(a+b+1)}\Bigg) \\
    \Var[K^{(0)}] & = & \E[(K^{(0)})^2] - (\E[K^{(0)}])^2 \nonumber \\
                  & = & \frac{N}{4}\Bigg(\frac{a(a+b)(a+2b+1) + Nab}{(a+b)^2(a+b+1)}\Bigg).
\end{IEEEeqnarray}
We have used the combinatorial identities (\ref{eqn:comb_id_1}) and (\ref{eqn:comb_id_2}) and also the fact that the second moment of a beta-binomial random variable $\kcount{m}{0} \sim \textrm{Beta-Binomial}(m, a, b)$ is given by $\frac{ma(m(a+1) + b)}{(a+b)(a+b+1)}$. The expressions for $\E[K^{(1)}]$ and $\Var[K^{(1)}]$ can be derived similarly.~\hfill\IEEEQED